\documentclass[10pt,onecolumn,draft]{IEEEtran}

\usepackage{amsmath}
\usepackage{amsthm}
\usepackage{amssymb}
\usepackage{mathtools}
\usepackage{graphicx}
\usepackage{rotating}
\usepackage{enumitem}
\usepackage{tcolorbox}
\usepackage{xcolor}
\usepackage{tikz}
\usepackage{bbold}
\usepackage{nicefrac}
\usepackage{dsfont}
\usepackage{cite}

\allowdisplaybreaks
\newsavebox{\tempbox}
\newcommand{\eric}[1]{\textcolor{violet}{\textbf{Comment:~#1}}}

\DeclareMathAlphabet\mbcf{OMS}{cmsy}{b}{n}

\DeclareMathOperator*{\argmin}{arg\,min}

\newcommand{\dx}{\mathrm{d}}

\newcommand{\mbf}[1]{\boldsymbol{#1}}
\newcommand{\mcf}[1]{\mathcal{#1}}
\newcommand{\idc}[2]{\mathbb{1}_{#2}\left({#1}\right) }

\newcommand{\com}[1]{\iftoggle{comments}{#1}{}}

\newtheorem{theorem}{Theorem}
\newtheorem{define}[theorem]{Definition}
\newtheorem{coding}[theorem]{Code Construction}
\newtheorem{codingmod}[theorem]{Code Modification}
\newtheorem{codingmodcor}[theorem]{Code Modification Corollary}

\newenvironment{proofsketch}{%
  \proof}{\endproof}

\newtheorem{lemma}[theorem]{Lemma}
\newtheorem{calc}[theorem]{Calculation}

\newtheorem{cor}[theorem]{Corollary}

\newtheorem{remark}[theorem]{Remark}
\newtheorem{remarkstar}[theorem]{Remark$^\star$}

\IEEEoverridecommandlockouts

\usepackage{etoolbox}

\providetoggle{comments}
\settoggle{comments}{false}

\begin{document}

\title{Keyless Authentication for AWGN Channels}

\author{Eric~Graves, 
        Allison~Beemer, 
        J\"{o}rg~Kliewer, 
        Oliver~Kosut, 
        and Paul~Yu~\IEEEmembership{}%
\thanks{A. Beemer is with the Department of Mathematics at the University of Wisconsin-Eau Claire, Eau Claire, WI, 54701;
J. Kliewer is with the Department of Electrical and Computer Engineering, New Jersey Institute of Technology, Newark,
NJ, 07103; 
O. Kosut is with the School of Electrical, Computer and Energy Engineering, Arizona State University, Tempe, AZ 85287; 
and E. Graves and P. Yu are with the Computer and Information Sciences Division, U.S. Army Research Laboratory, Adelphi, MD 20783.}%
\thanks{This research was sponsored by the Combat Capabilities Development Command Army Research Laboratory and was accomplished under Cooperative Agreement Number W911NF-17-2-0183. The views and conclusions contained in this document are those of the authors and should not be interpreted as representing the official policies, either expressed or implied, of the Combat Capabilities Development Command Army Research Laboratory or the U.S. Government. 
\com{So, if a statement like ``Trump's lack of understanding of exponential growth, and his deafness to the need for immediate testing has kinda fucked us pretty hard'' is expressed, while objectively true, it is not the official stance of the U.S. Government.} The U.S. Government is authorized to reproduce and distribute reprints for Government purposes not withstanding any copyright notation here on.}
}

\maketitle

\begin{abstract}
This work establishes that the physical layer can be used to perform information-theoretic authentication in additive white Gaussian noise channels, as long as the adversary is not omniscient.  
The model considered consists of an encoder, decoder, and adversary, where the adversary has access to the message, a non-causal noisy observation of the encoder's transmission, and unlimited transmission power, while the decoder observes a noisy version of the sum of the encoder and adversary's outputs.
A method to modify a generic existing channel code to enable authentication is presented. 
It is shown that this modification costs an asymptotically negligible amount of the coding rate, while still enabling authentication as long as the adversary's observation is not noiseless.
Also notable is that this modification is not (asymptotically) a function of the statistical characterization of the adversary's channel and furthermore no secret key is required, hence paving the way for a robust practical implementation. 
Using these results, the channel-authenticated 
\com{\eric{using ``authenticated'' instead of ``authentication.'' Authenticated capacity sounds to me more like information that has been authenticated, while authentication capacity sounds more like it's capacity to authenticate (i.e., probability of failure to authenticate). Personally feel there should be a distinction between the two, although I suspect they are equal.}}
capacity is calculated and shown to be equal to the non-adversarial channel capacity.
While this modular scheme is designed for use in the given channel model, it is applicable to a wide range of settings. 

\end{abstract}

\section{Introduction}

Authentication, or the act of verifying the identity of the source of information,  is a crucial aspect of security; especially in scenarios where the information leads to an observable action (e.g., calling in a missile strike or executing a stock market trade).
For information-theoretic authentication a decoder must be able to decode a message from the legitimate encoder while rejecting messages from a computationally-unbounded adversary.
More specifically, when an adversary is actively tampering with a message, the decoder only needs to output that the message is fake and does not need to output a message estimate.
Of course, a decoder declaration that the message is fake when it has not been tampered with is still considered an error. 
For the scenario here, we will consider information-theoretic authentication in the context of a classical communication system where the encoder, decoder, and adversary are connected by a noisy channel.
In such a context, information-theoretic authentication is generally achieved by exploiting a feature of the communication model that is unique between encoder and decoder and which the adversary cannot imitate.

In the existing literature two features are used: either exploiting the channel in such a way that the adversary cannot mimic a valid transmission, or by use of a secret key shared by encoder and decoder.
This work is classified in the former category, as we will not allow the encoder and decoder to share a secret key. 
For readers interested in secret key equipped information-theoretic authentication see Perazzone et al.~\cite{perazzone2020secret,graves2020secret} for an in-depth discussion on prior works and the best results to date.

In cases where no secret key is available, information-theoretic authentication can be obtained by exploiting (if possible) the uniqueness of the channel from the encoder to decoder.
This exploitation generally takes the form of choosing an encoder whose output when passed through the channel produces a set of observations that cannot be reliably reproduced by the adversary. 
Obviously then, the information that the adversary may act on and how they are allowed to act is crucial, as it determines how well they can mimic the legitimate encoder. 
Previous work~\cite{jiang2014keyless,jiang2015optimality,graves2016keyless,gungor2016basic,tu2018keyless,kosut2018authentication,beemer2019authentication,Sangwan19} on this topic is mainly differentiated by these particular formulation decisions. 
A few of these decisions to be made are (without vs. with): the allowance of joint transmission by adversary and encoder~\cite{jiang2014keyless,jiang2015optimality,tu2018keyless,gungor2016basic} vs. \cite{graves2016keyless,kosut2018authentication,beemer2019authentication,Sangwan19}, side information about the encoder's message at the adversary \cite{jiang2014keyless,jiang2015optimality,gungor2016basic,tu2018keyless,kosut2018authentication} vs. \cite{graves2016keyless,beemer2019authentication,Sangwan19}, and a noisy copy of the encoder's output at the adversary \cite{jiang2014keyless,jiang2015optimality,graves2016keyless,tu2018keyless,kosut2018authentication,Sangwan19} vs. \cite{gungor2016basic,beemer2019authentication}.
It is not surprising then that most of this work is similar in formulation, methodology, and results while still being diverse in terminology. {Our work here makes all three ``with'' allowances.}
For simplicity, we will broadly characterize~\cite{jiang2014keyless,jiang2015optimality,graves2016keyless,gungor2016basic,tu2018keyless,kosut2018authentication,beemer2019authentication,Sangwan19}.

To permit more formal discussion, let $p(y|x,v)$ be the conditional distribution of the decoder's input ($y$) given the encoder's output $(x)$ and the adversary's output $(v)$, and let $\mcf{Q}$ be the (model dependent) set of joint probability distributions for the encoder and adversary's output. Further assume that there is some symbol $\emptyset$ for a not-transmitting state.
As an example of how the formulation affects $\mcf{Q}$, if the encoder and adversary are not allowed simultaneous transmission then $\mcf{Q}$ will contain only distributions such that $q(x,v)>0$ only if $x= \emptyset$ or $v= \emptyset$. 
With this in mind, the previous literature divides the set of channels into sets based upon the property\footnote{Typically this property is denoted by an ``-able''-suffixed term, such as simulatable~\cite{jiang2014keyless,jiang2015optimality,gungor2016basic,tu2018keyless} (pace Maurer~\cite{maurer1994strong}), or overwritable~\cite{kosut2018authentication,Sangwan19}, $U/I$-overwritable~\cite{beemer2019authentication}. We abstained from naming the channel condition in~\cite{graves2016keyless}.} that
$$\min_{q \in \mcf{Q}} |p(y|x',\emptyset) - \sum_{x,v} p(y|x,v) q_{x'}(v,x)|_1 > 0  $$
for at least one encoder output $x$.
To understand how this property equips channels for information-theoretic authentication, view the distribution $w_{x'}(y) = \sum_{x,v} p(y|x,v) q(v,x)$ as the adversary's attempt to make the distribution at the input to the decoder close to the distribution that occurs when $x'$ is sent by the encoder and the adversary does not transmit, i.e. $p(y|x',\emptyset)$.
If $w_{x'}(y) \neq p(y|x',\emptyset)$ then the set of typical $k$-sequences of $y$ for this given $x'$ when the adversary is non-malicious (here denoted $\mcf{T}(x)$) has an exponentially decaying probability given the adversary is trying to imitate it, that is
$$w(\mcf{T}(x)) \leq 2^{-O(k \sqrt{ |w_{x'}(y) - p(y|x',\emptyset)  |_1 } ) }. $$
Hence a typical set detector can generally suffice to detect the manipulation. 


While not necessarily clear from the above discussion, the distribution of the encoder's output is important in determining $\mcf{Q}.$
That the capacity-achieving distribution at the output of the encoder also allows for authentication cannot be taken for granted.
Hence much of the previous work, specifically \cite{jiang2014keyless,jiang2015optimality,gungor2016basic,tu2018keyless,kosut2018authentication,Sangwan19,beemer2019structured}, opts for a two-code concatenated approach.
For this approach, one of the codes is a long (in terms of symbols) capacity-achieving code, while the other is a short low-rate code equipped with information-theoretic authentication.
Generally, the message is transmitted with the capacity achieving code, while a randomly generated number and a hash of the message and this randomly generated number are transmitted with the low-rate code that provides information-theoretic authentication. 
At this point it is important to note that the previous works that use this two stage approach do not give the adversary a noisy copy of the encoder output, and hence the adversary cannot possibly determine the randomly generated number used to construct the hash value.
Without knowledge of this number, and hence without the ability to modify the low-rate code, the adversary can at best hope that they will choose a message that when combined with the randomly generated number will result in the same hash value.

In this work, we will allow the adversary a noisy copy of the transmission; the addition of this extra channel is motivated by a common wireless communication scenario with an overwhelmingly strong adversary.
This overwhelmingly strong adversary will non-causally observe noisy versions of the encoder's output while also knowing the message that the encoder is transmitting. 
On the other end, the decoder will observe the superimposed transmissions of the adversary and encoder.
Both observations will be corrupted with independent \emph{additive white Gaussian noise} (AWGN), as is the tradition for first-order approximation to practical continuous-value channels dating back to Shannon~\cite{shannon1948mathematical}.

The allowance of non-causal observations at the adversary, in particular, is crucial for modeling since in practice it would be impossible to know the delay from the adversary to the encoder and from the adversary to the decoder.
Without knowing these delays, it would likewise be impossible to know how much of the encoder's output the adversary has observed and can therefore use in constructing their attack. 
Allowing a non-causal observation by the adversary thus corresponds to a worst-case scenario where the adversary has enough time to observe all of the encoder's output and then choose their own outputs accordingly.

Our desire to model realistic channels under extremely adverse conditions costs us both aspects of the traditional analysis. 
Indeed, recalling that a typical set detector is used to detect the manipulation, it is not surprising that the traditional analysis makes use of the fact that there are only a polynomial (in block length) number of different types.
By considering continuous channels, as opposed to~\cite{jiang2014keyless,jiang2015optimality,graves2016keyless,gungor2016basic,tu2018keyless,kosut2018authentication,beemer2019authentication,Sangwan19}, we can no longer take this approach. 
Furthermore, by allowing non-causal observations at the adversary, we are generally eliminating the option to use a two-code approach to obtain capacity.
To be sure, consider the case where the adversary has less noisy observations: here the adversary would be able to decode the random number used, and hence could determine the set of messages that would result in the same hash as the transmitted message.
Thus our choice of model, motivated by practical implementation, also requires a completely new approach to solve the problem.

Despite these adversarial advantages, our scheme will achieve information-theoretic authentication with the following notable features:
\begin{itemize}
    \item a construction based upon modifying almost any existing deterministic channel code;
    \item does not require a shared secret key or common randomness;
    \item will detect an adversary's manipulation as long as the adversary's observations of the encoder output's are not completely noiseless;
    \item achieves rates arbitrarily close to the non-adversarial channel capacity.
\end{itemize}
Thus, despite the austere channel model, our scheme still allows for a robust detection. 
We have specifically chosen the modification of arbitrarily given channel codes to provide a path forward for implementation. 
Our work allows researchers to concentrate on modifying existing codes already having good encoders and decoders, such as low-density parity-check codes, turbo codes, polar codes, or repetition codes.

We achieve the outcomes above by building on the insights of Graves et al.~\cite{graves2016keyless} and Beemer at al.~\cite{beemer2019authentication}, where authentication was enabled by introducing artificial noise at the output of the encoder.
In fact, both works show something even more surprising: there exist channels for which deterministic codes do not allow for information-theoretic authentication, but information-theoretic authentication can be enabled by adding artificial noise to the output of the encoder. 
For a simple example, consider a channel where the encoder can output $0$ or $1$, the adversary can output $-1$, $0$, and $1$, and where the decoder receives the sum of the two. 
Given any deterministic encoder $\mbf{x} : \mcf{M} \rightarrow  \{0,1\}^n$, if the adversary has knowledge of the transmitted message they may in turn choose their transmitted sequence as $\mbf{z}(M) =  \mbf{x}(a)-\mbf{x}(M)$  so that the decoder receives $$\mbf{y}(M) = \mbf{x}(M)  + \mbf{z}(M) = \mbf{x}(a),$$
which is indistinguishable from the case where the encoder sends $a\in \mcf{M}$  and the adversary does not interfere.
But now, instead consider a stochastic encoder constructed by simply taking a deterministic encoder and passing the output through a binary symmetric channel with positive crossover probability $p<\nicefrac{1}{2}$. 
Now, the probability of detection can be characterized as a function of the number of coordinates $i$ for which $z_i \neq 0.$ 
Indeed, assume that $z_i=1$, regardless of message the probability that the encoder outputs $1$ for the $i$-th coordinate is at least $p$, hence the probability that $y_i = 2$ is at least $p$, and $y_i = 2$ can only happen if the adversary is not sending $0$.
Thus it is easy to see that the probability of false authentication is at most $(1-p)^{|\{i | z_i \neq 0\}|}$.
This probability can be made arbitrarily small by starting with a well-chosen channel code. 
Hence a simple stochastic code gives us the ability to authenticate.
Of course our situation will be more complicated here because the adversary will have their own observation, but the premise remains the same.
Without complete knowledge of the encoder's output, the adversary's actions will result in decoder inputs that are not expected.

To take advantage of this insight, our code modification strategy consists of first adding carefully constructed message-dependent noise and then decimating the message set.
The message dependent noise is determined by a novel coding scheme that guarantees the adversary must always remove some of the noise added to the channel in order to forge a message.
As long as the adversary's observations themselves are noisy, the adversary will not be able to completely eliminate the message-dependent noise the encoder has added to the channel.
Thus, by detecting the presence of this noise the decoder can detect the adversary's presence.
This additional noise will guarantee that the adversary cannot modify a message to a specific message of their own choosing. 
From there, decimating the message set (a concept borrowed from Ahlswede and Dueck's local strong converse~\cite{bcgc}) extends this guarantee to ensure a small maximum probability of false authentication. 

To begin the formal treatment of this problem, the notation, model, and operational measures will be presented in Section~\ref{sec:nm}.
The results will be presented in Section~\ref{sec:results}, with many of the proofs being removed to the appendices for readability. Section \ref{sec:futures} includes a discussion of topics for further investigation, as well as a comparison of our scheme to secret key-based authentication schemes.
Conclusions are presented in the Section~\ref{sec:con}.

\section{Model and notation}\label{sec:nm}

\subsection{Notation}\label{sec:notation}

Uppercase letters will denote random variables, lowercase constants, and script sets.
In particular $\mcf{R}$ denotes the set of real numbers.

Bold font always denotes $n$-fold Cartesian products, with $n$ to be later defined as the block length of the code, and given $\mbf{x}$, $x_i $ is the $i$-th coordinate. 
In other words $\mbf{x} = \bigtimes_{i=1}^n x_i.$ 
While Cartesian products of random variables and constants may have unique coordinates, a set which is a Cartesian products of sets will not (i.e., $\mbcf{X} = \bigtimes_{i=1}^n \mcf{X}$).

Throughout the paper, $\mbf{G}_{\mbf{\rho}} = \bigtimes_{i=1}^n G_{\mbf{\rho},i}$ 
will be used to denote Cartesian product of $n$ independent Gaussian random variables with mean $0$ where the $i$-th coordinate has variance $\rho_i$.
When all the variances are equal, (i.e., $\mbf{\rho} = \bigtimes_{i=1}^n \rho$) just the single variance will be listed (i.e., $\mbf{G}_{\rho}$). 
Sometimes $\mbf{G}_{\mathrm{some~qualitative~value}}$ will be used in place of $\mbf{G}_{\mbf{\rho}_{\mathrm{some~qualitative~value}}}$ so that it is easier to specify the source of this randomness in the math. 
Finally all values of $G$, unless otherwise explicitly stated, should be assumed independent. 

All logarithms are natural, and the following functions will be used: 
\begin{align}
\mathbb{E}[ X] &= \int_{\mcf{R}} x f_{X}(x) \dx x 
    \notag \\ 
\mathbb{D}(X||Y) &= \int_{\mcf{R}} f_{X}(x) \log \frac{f_{X}(x)}{f_{Y}(x)} \dx x    
    \notag \\
\mathbb{D}_2(a||b) &= a \log \frac{a}{b} + (1-a) \log \frac{1-a}{1-b} 
    \notag \\
\mathbb{H}_2(a) &= -a \log  a - (1-a) \log (1-a) 
    \notag \\
\mathbb{I}_2\left( a || b \right)  & =
b \mathbb{D}_2(a||b) + (1-b) \mathbb{D}_2\left( b \frac{1-a}{1-b} \middle| \middle| b \right)
    \notag \\
&= \mathbb{H}_2(b) - b\mathbb{H}_2(a) - (1-b) \mathbb{H}_2\left( b \frac{1-a}{1-b} \right) 
    \notag\\ 
\mathbb{1}_{\mcf{A}}(b) &= \begin{cases} 1 &\text{if } b \in \mcf{A}    \\ 0 &\text{else} \end{cases}
    \notag \\
\Phi(x) &= \int_{-\infty}^x \frac{1}{\sqrt{2 \pi}} e^{- \frac{t^2}{2}} \dx t 
    \notag
\end{align}
where $ f_X$ is used to denote the probability density function of $X$.
Furthermore, we will use $\left( \begin{matrix} \mcf{A} \\ b \end{matrix} \right)$ to denote the set of all $b$-element subsets of $\mcf{A}.$ 
For instance $$\left( \begin{matrix} \{1,2,3\} \\ 2 \end{matrix} \right) = \left\{ \{1,2\}, \{1,3\}, \{2,3\} \right\}.$$

\subsection{Model}\label{sec:model}

\tikzstyle{circ} = [draw, fill=white, circle, node distance=1cm]

\tikzstyle{block} = [draw, fill=white, rectangle, 
    minimum height=30pt, minimum width=30pt, text centered]

\tikzstyle{bigblock} = [draw, fill=white, rectangle, 
    minimum height=100pt, minimum width=20pt, text centered]
    
\begin{figure*}[t!]    
    \begin{center}
\begin{tikzpicture}[thick, every node/.style={transform shape}]

\node[block] (enc) at (0,0) {$\begin{array}{c} \text{Encoder} \\  \mbf{X}(M) \end{array}$};
\node[circ] (add1) at (2,-.75) {$+$};
\node[circ] (add2) at (6,0) {$+$};

\node[block] (dec) at (4,-1.7){$\begin{array}{c} \text{Adversary} \\  \mbf{Z}(\mbf{V},M)  \end{array}$};


\node[block] (dec2) at (9,0){$\begin{array}{c} \text{Decoder} \\  \hat m( \mbf{Y} )  \end{array}$};

\draw[->,thick,dashed] (3.5,-.75) node[above]{$\mbf{G}_{\mathrm{Adv}}$}   -- (add1.east) ;
\draw[->,thick] (6,.75) node[right]{$\mbf{G}_{\mathrm{Dec}}$}   -- (add2.north) ;

\draw[->,thick] (dec.east) -- (6,-1.7) node[below] {$\mbf{Z}$} -- (add2.south);
\draw[->,thick] (enc.east) -- (add2.west);
\draw[->,thick] (add2.east) -- (6.75,0) node[above]{$\mbf{Y}$} -- (dec2.west);
\draw[->,thick] (dec2.east) -- (11,0) node[above]{$\hat M$};

\draw[->,thick] (-2,0) node[above] {$M$} -- (enc.west);

\draw[->,thick,dashed] (2,0) node[above] {$\mbf{X}$} -- (add1.north);
\draw[->,thick,dashed] (add1.south) -- (2,-1.45) node[left] {$\mbf{V}$} -- (2.96,-1.45) ;
\draw[->,thick,dashed] (-1.5,0) -- (-1.5,-1.95) -- (2.96,-1.95);



\end{tikzpicture}

\end{center}
\caption{Channel with encoder $\mbf{X}:\mcf{M} \rightarrow \mbcf{R}$ and decoder $\hat m: \mbcf{R} \rightarrow \mcf{M} \cup \{!\}$, where $\mbf{G}_{\mathrm{Dec}} \sim \text{Gaussian}(0,\rho_{\mathrm{Dec}})$ and $\mbf{G}_{\mathrm{Adv}} \sim \text{Gaussian}(0,\rho_{\mathrm{Adv}})$. The dashed lines represent non-causal links.} \label{fig:chan}
\end{figure*}
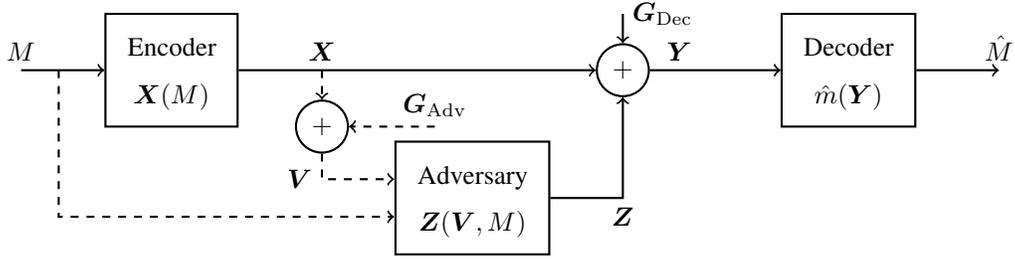

The model for communications (pictured in Figure~\ref{fig:chan}) studied here consists of three entities: an encoder, decoder, and an adversary.
\com{\eric{Used ``in this model'' instead of deleting ``here'' to avoid the same word starting back to back sentences}} In this model, the encoder is tasked with sending a message $M$ to the decoder, where the message is assumed to be uniform\footnote{The distribution of this message will not play a role in the results.} over $\mcf{M}$.
To do this the encoder will map the message to an $n$-symbol sequence $\mbf{X}(M)$ and send it across the communications channel. It is important to note that the code is allowed to be a random function of the message.

When the encoder sends its codeword, the adversary will receive a non-causal noisy copy of the $n$-symbol sequence 
$$\mbf{V} = \mbf{X}(M) + \mbf{G}_{\mathrm{Adv}},$$
where $\rho_{\mathrm{Adv}} \in (0,\infty)$ represents variance of the adversary's noise.
Using this received information, the adversary will craft their own $n$-symbol sequence to inject into the channel.
In general this function will be modeled by $\mbf{Z} : \mbcf{R} \bigtimes \mcf{M} \rightarrow \mbcf{R}$.
Towards discussion purposes, it can be generally assumed that the adversary chooses this function optimally: that is, to minimize the performance metrics of the system. 

On the other hand, the decoder will receive a noisy copy of the combination of $n$-symbol sequences sent by encoder and adversary,
$$\mbf{Y} = \mbf{X}(M) + \mbf{Z}(\mbf{V},M) + \mbf{G}_{\mathrm{Dec}},$$
where $\rho_{\mathrm{Dec}}\in (0,\infty)$ represents the noise variance at the decoder.
From there, the decoder will attempt to estimate the message the encoder sent as $\hat m(\mbf{Y})$ or will output $\mbf{!}$ to indicate that the adversary has altered the message.

\subsection{Operational Parameters} \label{sec:opparam}

The objective of this work is to construct a good code for authenticated communications.
\begin{define}\textbf{(Code)} 
\com{\eric{Stochastic code in the sense that the output is stochastic, not in the sense that a code is randomly chosen before hand. Although, I feel that the difference between those two is more philosophical than mathematical. Like, if a coin is flipped but you do not know the results, is it still random? Regardless, the terminology can be changed to ``random code,'' but this creates problems with code construction.}} A \emph{code} is a set of paired functions $\mbf{X} : \mcf{M} \rightarrow \mbcf{R}$, $\hat m : \mbcf{R} \rightarrow \mcf{M} \cup \{\mbf{!}\}$ representing the \emph{encoder} and \emph{decoder} respectively. The symbol $\mbf{!}$ specifically represents the case that the decoder labels the observation as not authentic. 
\end{define}
\begin{remark} 
A code not designed for authenticated communications can be considered as a special case where $\hat m(\mbf{y}) \neq \mbf{!}$ for all $\mbf{y} \in \mbcf{R}.$
\end{remark}
\begin{remark}
Codes are assumed to have block length (number of symbols output) $n$, unless otherwise stated.
\end{remark}

Codes will be measured by the rate at which they can send information, the power required to do so, the reliability with which information is decoded when there is no adversarial interference, and the likelihood the adversary can manipulate the decoder into accepting a false message. 
Formal definitions for the first three follow.
\begin{define}\label{def:rate}\textbf{(Rate)} The \emph{rate} of a code $\mcf{H} = (\mbf{X}, \hat m)$ is 
$$r_{\mcf{H}} = \frac{1}{n} \log |\mcf{M}|.$$
\end{define}
\begin{define}\label{def:power}\textbf{(Power Constraint)} The \emph{power constraint} of a code $\mcf{H} = (\mbf{X}, \hat m)$ is 
$$\omega_{\mcf{H}} = \max_{m \in \mcf{M}}  \sum_{i=1}^n \frac{1}{n} \mathbb{E}\left[ X_i^2(m) \right] .$$
\end{define}
\begin{define}\label{def:prerr}\textbf{(Error Probability)}
For code $\mcf{H} = (\mbf{X}$, $\hat m$) the \emph{arithmetic average error probability} at noise variance $\rho_{\mathrm{Dec}} \in (0,\infty)$ is
$$
\varepsilon_{\mcf{H}}(\rho_{\mathrm{Dec}}) = \sum_{m \in \mcf{M}} \frac{1}{|\mcf{M}|} \Pr \left( \hat m ( \mbf{X}(m) + \mbf{G}_{\mathrm{Dec}} )  \neq m  \right) .
$$
\end{define}
\noindent Note that the error probability is indeed a measure of reliability when not under adversarial influence, since if $\mbf{Z}(\mbf{V},M) = \mbf{0}$ then  $\mbf{Y} = \mbf{X}(M) + \mbf{G}_{\mathrm{Dec}}$.

Two measures of the adversary's ability to interfere will be considered. 
The weaker of these two measures considers the adversary's ability to have the decoder accept a specific message.
\begin{define}\label{def:tfa}\textbf{(Maximum Probability of Targeted False Authentication)}
The \emph{maximum probability of targeted false authentication} for code $\mcf{H}= (\mbf{X}, \hat m)$ with decoder noise variance $\rho_{\mathrm{Dec}} \in  ( 0 , \infty)$ and adversary noise variance $\rho_{\mathrm{Adv}} \in (0,\infty)$ is
\begin{align}
&\alpha^*_{\mcf{H}}(\rho_{\mathrm{Dec}}, \rho_{\mathrm{Adv}}) =
    \notag \\
&\sup_{\substack{\mbf{Z}:\mbcf{R} \bigtimes \mcf{M} \rightarrow \mbcf{R} \\ a \in \mcf{M} \\ b \in \mcf{M} \setminus \{a\}} } \Pr \left( \hat m(\mbf{X}(a) + \mbf{G}_{\mathrm{Dec}} + \mbf{Z}(\mbf{V},a) ) = b   \right) 
    \notag ,
\end{align}
where 
$$\mbf{V} = \mbf{X}(a) + \mbf{G}_{\mathrm{Adv}}.$$
\end{define}
A small probability of targeted false authentication does not guarantee the decoder will not output a false message, instead it guarantees that the adversary cannot \textit{choose} which message it is.
This weaker metric will only play a brief role in this study, with the main goal being to obtain codes which measure favorably under the following, stronger metric. 
\begin{define}\label{def:pa}\textbf{(Maximum Probability of False Authentication)}
The \emph{maximum probability of false authentication} for code $\mcf{H}= (\mbf{X},\hat m)$ with decoder noise variance $\rho_{\mathrm{Dec}}$ and adversary noise variance $ \rho_{\mathrm{Adv}}$ is
\begin{align}
&\alpha_{\mcf{H}}(\rho_{\mathrm{Dec}}, \rho_{\mathrm{Adv}}) =
    \notag \\
&\sup_{\substack{\mbf{Z}:\mbcf{R} \bigtimes \mcf{M} \rightarrow \mbcf{R} \\ a \in \mcf{M}} }  \Pr \left( \hat m(\mbf{X}(a) + \mbf{G}_{\mathrm{Dec}} + \mbf{Z}(\mbf{V},a) ) \notin \{ a,\mbf{!}\}   \right) ,
    \notag 
\end{align}
where 
$$\mbf{V} = \mbf{X}(a) + \mbf{G}_{\mathrm{Adv}}.$$
\end{define}

Unlike the targeted false authentication probability, a vanishing probability of false authentication does asymptotically guarantee the decoder will not output a false message in the presence of an adversary.

\begin{remark}
To better understand the relationship between the two metrics observe that
\begin{align}
&\Pr \left( \hat m(\mbf{X}(a) + \mbf{G}_{\mathrm{Dec}} + \mbf{Z}(\mbf{V},a) ) \notin \{ a,\mbf{!}\}   \right)
    \notag \\ &\quad  
    = 
 \sum_{b \in \mcf{M}\setminus\{a,\mbf{!}\}} \Pr \left( \hat m(\mbf{X}(a) + \mbf{G}_{\mathrm{Dec}} + \mbf{Z}(\mbf{V},a) ) = b  \right),
 \notag
\end{align}
from which it is clear that having a small maximum probability of targeted false authentication does not guarantee a small maximum probability of false authentication, but a small maximum probability of false authentication does guarantee a small maximum probability of targeted false authentication.
\end{remark}
\begin{remarkstar}\label{remark:iwonderifthiswill work}
Readers familiar with information-theoretic authentication literature\footnote{Primarily, the information-theoretic authentication literature whose genesis is Simmons~\cite{Auth}; most of these works are secret key-based {which marks a striking difference from our work here}.} may be wondering why we have not defined the impersonation attack.
For those unfamiliar, an impersonation attack is one where the adversary does not wait for the encoder to produce an output, but directly sends a value to the decoder. 
If we were to formally define this as an operational measure it would be 
\begin{align}
&\sup_{\mbf{Z}:\mbcf{R} \bigtimes \mcf{M} \rightarrow \mbcf{R}  }  \Pr \left(  \mbf{G}_{\mathrm{Dec}} + \mbf{Z}(\mbf{G}_{\mathrm{Adv}},\emptyset)  \notin \{ \emptyset, \mbf{!}\}   \right) .
    \notag 
\end{align}

We do not feel the need to define this metric separately, since it can already be accounted for in the definition of the encoder and decoder. 
That is, in $\mcf{M}$ we may assume that there is a special symbol (call it $\emptyset$) that corresponds to the case where the encoder has no message to transmit. 
If, for instance, $\mbf{X}(\emptyset) = \mbf{0}$ then clearly
\begin{align}
&\alpha_{\mcf{H}}(\rho_{\mathrm{Dec}}, \rho_{\mathrm{Adv}}) 
    \notag \\ &\quad 
    \geq
\sup_{\mbf{Z}:\mbcf{R} \bigtimes \mcf{M} \rightarrow \mbcf{R} }  \Pr \left( \hat m( \mbf{G}_{\mathrm{Dec}} + \mbf{Z}(\mbf{G}_{\mathrm{Adv}},\emptyset) ) \notin \{ \emptyset ,\mbf{!}\}   \right) ,
    \notag 
\end{align}
and hence our formulation (for appropriately defined codes) already encompasses impersonation attacks.    

This observation will have an important consequence in the context of our results. 
In preview of this, in order to ensure authentication, our results will require the encoder still output low levels of additive white Gaussian noise when it has no message to transmit. 
Clearly the assumption of such a possibility will be model-dependent and would not be valid for a situation like a wired channel where the channel may be physically severed. 
On the other hand, severing a link would be very difficult in a wireless environment and hence an eternally active encoder in our model is justifiable.
\end{remarkstar}

One of the primary goals of our work will be to characterize the authenticated capacity. 
Intuitively, the authenticated capacity is the maximum rate possible under a given power constraint and the requirement that the probability of error and maximum probability of false authentication converge to zero. 
In order to present the exact definition, the notation 
\com{\eric{Do not understand comment. Defining the message set size as a function of the block length allows for a sequence of codes with near constant rate.}} 
$\mbf{X}_{(n)}: \mcf{M}_{(n)} \rightarrow \mbcf{R}_{(n)}, \hat m_{(n)} : \mbcf{R}_{(n)} \rightarrow \mcf{M}_{(n)} \cup \{\mbf{!}\}$ will be used to denote codes with block length $n$.

\begin{define}\label{def:cap}\textbf{(Authenticated capacity)}
The \emph{authenticated channel capacity} is 
%
%
\begin{align}
&c(\rho, \rho_{\mathrm{Dec}},\rho_{\mathrm{Adv}}) =
    \notag \\
& \sup \left\{ r \in \mcf{R} \middle| 
\begin{array}{rl} \exists \mcf{H}_{(n)} =  \{ \mbf{X}_{(n)}, \hat m_{(n)} \}_{n=1}^\infty & \\
\text{ such that }&  \\
\displaystyle \limsup_{n \rightarrow \infty} \omega_{\mcf{H}_{(n)}}  &\leq  \rho \\
\displaystyle\liminf_{n \rightarrow \infty} r_{\mcf{H}_{(n)}} &\geq r \\
\displaystyle\limsup_{n \rightarrow \infty} \varepsilon_{\mcf{H}_{(n)}}(\rho_{\mathrm{Dec}}) &= 0 \\
\displaystyle\limsup_{n \rightarrow \infty} \alpha_{\mcf{H}_{(n)}}(\rho_{\mathrm{Dec}},\rho_{\mathrm{Adv}}) &= 0 
\end{array} \right\}.
    \notag
\end{align}
\end{define}
\begin{remark}
If the authentication requirement were removed, the capacity would be 
$$\frac{1}{2} \log \left( 1 + \frac{\rho}{\rho_{\mathrm{Dec}}} \right) $$
following from Shannon~\cite{shannon1948mathematical}. 
Indeed, removing the authentication measure leaves the operational definitions for point-to-point communications over an AWGN channel without an adversary, as is to be expected.
\end{remark}

On a final note, as mentioned in the introduction, Section~\ref{sec:results}'s code construction results will be presented in terms of a given channel code.
These results will, however, require the initial channel code be deterministic, i.e., the encoder output is not random given the message. 

\section{Results}\label{sec:results}

In this section we will build a number of consecutive results which lead to the conclusion that information-theoretic authentication is possible in AWGN channels without the need for a secret key.
Not only this, but we will also show that existing codes can be equipped with information-theoretic authentication at a small cost to the rate, power, and error probability of the code. 
This is achieved by modifying the codes with two complementary modifications.
The first of these modifications will (literally) add to the encoder's output a type of code which enables detection of targeted authentication attacks. 
The second modification of eliminating messages will then extend this to cover all attacks. 
With these results in hand, we show that the costs asymptotically vanish while the ability to detect manipulation remains; furthermore, we show that this is true regardless of the difference between the adversary and decoder's noise variance, instead only requiring that the noise variance at the adversary be non-zero.

In pursuit of a modular scheme, we begin by constructing a new type of code, termed an \textit{overlay code}.
Conceptually, these codes are used to control the amount of a persistent\footnote{By ``persistent'' we mean that it is difficult to remove.} resource added to each transmission symbol for each message.
The overlay code guarantees that a portion of this persistent resource must be removed by the adversary before they can falsify a message.
If the adversary is unable to remove the persistent resource, then its presence can be used by the decoder to detect the intrusion.
For the given channel model, the persistent resource will take the form of Gaussian noise, and the adversary will have to attempt noise cancellation in order to remove the persistent resource's presence. 

Before introducing overlay codes in Definition~\ref{def:overlay}, it will be helpful to introduce the intuition behind their conception. 
These codes are structured to enable basic statistical testing practices to detect the overabundance of the persistent resource. 
This is done by first limiting to a discrete set the possible levels of persistent resource added per symbol.
All symbols that have a given amount of persistent resource (e.g., all symbols which have had half of the maximum amount of resource added) can be thought of as the ``test sets'' since these sets will eventually form the sets over which we perform hypothesis testing {in order to determine the presence of an adversary}.
The most important property of the overlay code is that for any given message and any alternative message, one of the test sets for the given message will correspond to symbols whose persistent resource level is always less than or equal to (with a certain amount guaranteed to be strictly less than) the persistent resource level of the alternative message.
Consider this set-up in the context of authentication, where the alternative message represents the actual transmitted message and the given message the one produced by the decoder.
In this case, one of the test set for the given (decoded) message will correspond to a set of symbols for which the encoder added more of the persistent resource for the alternative (transmitted) message.
If the adversary cannot remove this resource efficiently enough, then its presence can be used to detect the message is false.
We now define the overlay code.

\begin{define} \label{def:overlay}
Given finite set $\mcf{K} \subset[0,1)$, and $\mcf{\tilde K} = \mcf{K} \cup 1,$ positive real number $r$, and $\gamma \in \left( \frac{1}{2}, 1 \right)$, a function $\mbf{f}: \mcf{M} \rightarrow \mbcf{\tilde K}$ is a \emph{$(r,\mcf{K} ,\gamma)$-overlay code} when
\begin{itemize}
\item  $$\frac{1}{n}\log |\mcf{M}|\geq r;$$
\item $$  \sum_{i=1}^n \idc{ f_i(m)}{\{ k \} }  =  \ell := \left \lfloor \frac{n}{|\mcf{\tilde K}|} \right \rfloor 
$$
for all  $m \in \mcf{M}$  and $k \in \mcf{K}$;
\item and for each distinct $m, m' \in \mcf{M}$ there exists a $k \in \mcf{K}$ such that 
$$ \sum_{i=1}^n \idc{ f_i(m)}{\{k\}} \idc{ f_i(m')}{\{ k\}}  \leq  \gamma  \ell $$
and for all $j \in \mcf{K}$ such that $j < k$
$$ \sum_{i=1}^n \idc{ f_i(m)}{ \{k\} } \idc{ f_i(m')}{\{j\}}  =   0 .$$
\end{itemize}

\emph{Uniform overlay codes} are overlay codes with $\mcf{K} = \left\{0, |\mcf{\tilde K}|^{-1}, \dots, 1 - |\mcf{\tilde K}|^{-1} \right\}$.
\end{define}

\begin{remark}
For the remainder of the paper, let $\mcf{\tilde K} := \mcf{K} \cup 1$ and $\ell:= \left \lfloor \frac{n}{|\mcf{\tilde K}|} \right \rfloor.$
\end{remark}
\begin{remark}
If $\mbf{f}$ is an $(r,\mcf{K},\gamma)$-overlay code, then for each $\mcf{\tilde M}\subset \mcf{M}$ the function $\mbf{\tilde f} : \mcf{\tilde M} \rightarrow \mbcf{R}$ defined by $\mbf{\tilde f}(m) = \mbf{ f}(m)$ is a $(\frac{1}{n} \log |\mcf{\tilde M}|,\mcf{K},\gamma)$-overlay code.  
\end{remark}
\com{\eric{$\mcf{M}$ and $n$ are defined in the model section as the set of messages $M$ is distributed over, and the number of symbols transmitted respectively. Also, notation defined earlier is that $\mbf{x} = \times_{i=1}^n x_i $, which should imply $\mbf{f} = \times_{i=1}^n f_i$. I would be ok with changing it to $f(m)_i$, which would be taking the $i$-th coordinate of the output of $\mbf{f}(m).$} }
\begin{remark}
It would certainly be possible to define overlay codes to allow a non-uniform number of symbols per persistent resource level (less than the maximum). That we did not do so is merely for the sake of simplicity. 
\end{remark}

Note, fewer resource levels $|\mcf{\tilde K}|$ implies more symbols share each level, hence fewer levels implies that there are more symbols to test per set.
Obviously though, fewer resource levels also means fewer unique output sequences for the overlay code, hence the overlay code will support fewer messages. 
To quickly see this, observe that if $\mcf{\tilde K}$ consisted of two elements, then there would be at most $2^n$ different possible code combinations.


The existence of overlay codes should not be taken for granted a priori.
For instance, consider a traditional random coding argument where for each message the encoder outputs are chosen at random from a predefined distribution. 
For any two messages $a$ and $b$, let $F_i(a)$ and $F_i(b)$ denote the randomly chosen value of $i$-th coordinate resource level for messages $a$ and $b$. 
Observe that $ \Pr \left( F_i(a) < F_i(b) \right) = \frac{1 - \sum_{k \in \mcf{\tilde K}} \Pr\left( F_i(a) = k \right)^2 }{2}$. 
Thus for any choice of distribution other than a deterministic one, $\Pr \left( F_i(a) < F_i(b) \right)>0$, and hence when rate $r> - n^{-1}\log \Pr \left( F_i(a) < F_i(b) \right)$ this construction will (with near certainty) produce a code such that for every message $a$, there exists a message $b$ whose resource levels are always greater than or equal to $a$'s.
Increasing the size of $\mcf{K}$ would exacerbate this problem.
Nevertheless, overlay codes do exist given certain conditions outlined in Theorem~\ref{thm:codevid19} and Corollary~\ref{cor:wedontwantthemtogetofftheshipbecausethatmaydoubleournumbers}.

\begin{theorem}\label{thm:codevid19}
For any positive real number $r$, finite $\mcf{K}\subset [0,1)$, and $\gamma \in \left( \frac{1}{2}, 1 \right)$  such that
 $$r \leq \frac{1}{n} \sum_{k \in \mcf{K}} n_k \left| \mathbb{I}_2\left(\gamma  \middle|\middle| \frac{\ell}{n_k} \right) - \frac{4}{3n_k} - \frac{2}{n_k}\log  n_k \sqrt{\ell} \right|^+,$$
where $n_k = n - \ell |\{ j \in \mcf{K}|j< k\}|$, there exists a $(r,\mcf{K},\gamma)$-overlay code.
\end{theorem}
\begin{cor}\label{cor:wedontwantthemtogetofftheshipbecausethatmaydoubleournumbers}
For all  $\gamma \in \left( \frac{1}{2}, 1 \right)$ and finite $\mcf{K}\subset [0,1)$, if positive number
$$r < \gamma \log (|\mcf{\tilde K}|)   - \gamma - \mathbb{H}_2(\gamma) , $$
then for large enough $n$ there exists a $(r,\mcf{K},\gamma)$-overlay code.
\end{cor}
\begin{proofsketch} 
The full proofs of Theorem~\ref{thm:codevid19} and Corollary~\ref{cor:wedontwantthemtogetofftheshipbecausethatmaydoubleournumbers} can be found in Appendix~\ref{app:codevid19}. Also to be found in Appendix~\ref{app:codevid19} is a detailed example of the overlay code construction.

We prove the theorem using an iterated random coding procedure. 
First  we represent $\mcf{M}$ as {in bijection with} a product of smaller sets, that is
$\mcf{M} = \bigtimes_{i\in \{1,\dots,|\mcf{K}|\}} \mcf{M}_{i} .$
Next, independently for each $m_1 \in \mcf{M}_1$ we randomly select an $\ell$-coordinate subset out of the total $n$ coordinates. 
These $\ell$ coordinates are those for which the overlay code outputs the smallest resource concentration (i.e., the minimum value in $\mcf{K}$).
This process is repeated for all $(m_1, m_2) \in \mcf{M}_1 \times \mcf{M}_2$, with the difference being that the set of $n-\ell$ coordinates not selected for $m_1$ is used {for selection of coordinates of the second-smallest resource concentration}.
This process of removing the selected coordinates and then randomly selecting a new set of coordinates is repeated until there are fewer than $\ell$ coordinates remaining; at this point the remaining coordinates are assigned an overlay output of $1$.

From this process, the resulting form of Theorem~\ref{thm:codevid19} should be clear. 
The summand over each $k \in \mcf{K}$ is simply the maximum rate at which our analysis can guarantee that two messages {match on} at most $\gamma \ell$-chosen coordinates.

To see why this method works, consider the following.
For any two messages $m,m' \in \mcf{M}$ there exist representations ($m_1,\ldots, m_{|\mcf{K}|})$ and $(m_1',\ldots, m_{|\mcf{K}|}')$ respectively. 
Clearly, there exists a smallest value $j\in \{1,\dots, |\mcf{K}|\}$ such that $m_j \neq m_j'.$
Now, for $m$ and $m'$ the overlay code coordinates corresponding to the $1$st through $(j-1)$th resource levels will be equal since $(m_1,\ldots, m_{j-1}) = (m_{1}',\ldots, m_{j-1}')$.
For the $j$th level though, the two messages will have different coordinates.
Furthermore, whenever the output overlay concentration for message $m$ is equal to the $j$th level, the concentration for message $m'$ must be greater than or equal to the $j$th level since all coordinates for resource levels less than that level are shared.
Using the appropriate random coding techniques, we can then guarantee a certain percentage of coordinates that do not share a level for $m_{j}$ and $m_{j}'.$

\end{proofsketch}

\begin{remark}
Of extreme importance here is that for a fixed rate $r$ and fixed $\gamma$, there is a fixed $|\mcf{K}|$ that guarantees the existence of a overlay code for large enough $n$. 
Thus, the value of $|\mcf{K}|$ should be intuitively viewed as a constant when dealing with asymptotic results.
\end{remark}

\begin{remark}
We will not be concerned with choosing the optimal values for inclusion in $\mcf{K}$ in this paper. 
This is primarily because the optimal values will depend on the adversary's noise variance, and we wish to have our code construction be independent of this knowledge.  
We will return to this discussion in Section~\ref{sec:futures}.
\end{remark}


Given the existence of overlay codes, we now go about applying them to arbitrary codes to enable authentication.
Importantly, a secret key is not necessary in this application, since authentication is enabled by the persistence of the resource added.
For our communication model, the persistent resource is additive Gaussian noise. 
Our code modification will make use of the overlay code to determine the variance of the Gaussian noise added to the encoder's output.
For primarily clerical reasons, another message-dependent signal, $\mbf{t}(M)$, will also be added to the output of the encoder. 
We strongly suspect it is not necessary for most practical codes, although it is necessary for a result that is agnostic of the original code.

\begin{codingmod} \label{code:addnoise}

~\\
Suppose 
\begin{itemize} 
\item a deterministic code $\mbf{x}: \mcf{M} \rightarrow \mbcf{R}$, $\hat m : \mbcf{R} \rightarrow \mcf{M}$, 
\item injection noise power $\rho_{\Delta} \in (0,\infty)$, 
\item tolerance $\delta\in (0,1)$, and
\item an $\left( \frac{1}{n} \log |\mcf{M}|, \mcf{K}, \gamma \right)$-overlay code $\mbf{f}:\mcf{M} \rightarrow \mbcf{\tilde K}$, for some finite $\mcf{K} \subset [0,1)$ and $\gamma \in \left(\frac{1}{2},1\right)$,
\end{itemize}
are given.



Independently for each $m \in \mcf{M}$ and $i\in\{1,\dots,n\}$ randomly choose $t_i(m)\in \mcf{R}$ according to a Gaussian distribution with mean $0$ and variance $(1-f_i^2(m)) \rho_{\Delta}$. 
Define the modified encoder $\mbf{ X}': \mcf{M} \rightarrow \mbcf{R}$ by
$$\mbf{X}'(M) = \mbf{x}(M) + \mbf{t}(M)+ \mbf{f}(M) \cdot \mbf{G}_{\Delta},$$
where $\cdot$ is the coordinate-wise product and $\mbf{G}_{\Delta} = \mbf{G}_{\rho_{\Delta}}$.
Define the modified decoder $\hat m': \mbcf{R}\rightarrow \mcf{M} \cup \{ \mbf{!}\}$ by
\begin{align} 
&\hat m' (\mbf{y})  
    \notag \\ &=
\begin{cases}
\hat m(\mbf{y}) & \text{if } \forall k \in \mcf{K}  \\
& \displaystyle \sum_{i \in \mcf{I}_k} \frac{[y_i -t_i(\hat m(\mbf{y}))- x_i(\hat m (\mbf{y}))]^2}{k^2 \rho_{\Delta} + \rho_{\mathrm{Dec}}} \leq \ell ( 1+ \delta)  \\
\mbf{!} &\text{else}
\end{cases}, \notag 
\end{align}
where $\mcf{I}_k$ is the set of coordinates $i$ such that $f_i(\hat m (\mbf{y}) ) = k.$

The resulting modified code is defined by $\mbf{X}',\hat m'.$

\end{codingmod}

\begin{remarkstar}\label{remark:alwayson}
Recall Remark~\ref{remark:iwonderifthiswill work} from Section~\ref{sec:model}.
In this remark we noted how our single metric could handle both impersonation attacks and substitution (or inference) attacks by assuming that the code had a message that corresponded to a ``not transmitting'' state.
The application of this code modification must also apply to this ``not transmitting'' state.
In other words, Code Modification \ref{code:addnoise} requires that the encoder still send a low level noise when there is no message to transmit.  
\end{remarkstar}

\begin{remark}
Note the modified decoder is the original decoder with the extra requirement that 
$$ \sum_{i \in \mcf{I}_k} \frac{[y_i -t_i(\hat m(\mbf{y}))- x_i(\hat m (\mbf{y}))]^2}{k^2 \rho_{\Delta} + \rho_{\mathrm{Dec}}} \leq \ell ( 1+ \delta) $$
for all $k \in \mcf{K}$.
In this sense, the modified decoder can be viewed as first using the original decoder to decode the message, and then checking for manipulation by ensuring that the extra requirement is met. 
For reference purposes, we shall adopt this two-stage decoder view, and refer to the checking of the extra requirement as the \emph{detector}.
\end{remark}

To see why Code Modification~\ref{code:addnoise} provides a small probability of \textit{targeted} false authentication, consider the steps an adversary would have to perform in order to fool the decoder into authenticating a particular message.
First, the adversary, given their a priori knowledge of the message and codebook, would subtract out the output of the unmodified encoder for the transmitted message as well as the $\mbf{t}$ term.
Next they would add in the unmodified encoder's output and the $\mbf{t}$ term for the alternative message they wished the decoder to accept. 
Finally, the adversary would try to ensure that the correct amount of noise is applied to the correct symbols {so as to avoid detection}.

But, while the adversary knows the variance of the encoder-added noise per symbol, they will not know the exact value of this added noise since their measurement is itself noisy. 
As the injected noise power becomes smaller, the variance of the adversary's estimate will become increasingly large relative to the encoder-added noise's own variance. 
Eventually, the adversary's estimate will be so poor that if the adversary tries to cancel out the added noise the resulting variance would not be significantly less than that of the encoder-added noise alone. 
Thus the scheme protects against any message being forged into a different particular message, since this different message will be guaranteed to have a set of coordinates that have less noise variance per symbol than the adversary can manage.
Later we will extend this scheme using Code Modification~\ref{code:decimate}/Code Modification Corollary~\ref{code:decimate_alt} to protect against all types of attacks.

While this does provide a form of information-theoretic authentication, adding noise to the output of the encoder will degrade the signal-to-noise ratio.
In turn, this decrease in the signal-to-noise ratio will reduce the maximum achievable rate or, alternatively, increase the probability of decoding error.
Our analysis favors the increase in the probability of error.
Additionally, the increase in noise will increase the power needed by the encoder.
But, as Theorem~\ref{thm:1stcode}/Corollary~\ref{cor:1stcode} formally shows, these costs can vanish while still allowing detection of \emph{targeted} authentication attacks.

\begin{cor}[Theorem~\ref{thm:1stcode}]\label{cor:1stcode}
Given injection noise power $\rho_{\Delta} = o(1)  $ and tolerance $\delta =  o(\rho_{\Delta})$,  then for all 
\begin{itemize} 
\item deterministic codes $\mcf{H} = (\mbf{x}: \mcf{M} \rightarrow \mbcf{R}$, $\hat m : \mbcf{R} \rightarrow \mcf{M})$, 
\item $\left( r_{\mcf{H}} , \mcf{K}, \gamma \right)$-uniform overlay code $\mbf{f}: \mcf{M} \rightarrow \mbcf{\tilde K}$ for any $\gamma \in (1/2, 1)$ and viable $|\mcf{K}|$,
\item and large enough $n$ ,
\end{itemize}
Code Modification~\ref{code:addnoise} yields with high probability a code $\mcf{J} = (\mbf{X}' : \mcf{M} \rightarrow \mbcf{R}$, $\hat m': \mbcf{R} \rightarrow \mcf{M} \cup \{\mbf{!} \} )$ such that   
\begin{align}
r_{\mcf{J}} &= r_{\mcf{H}} 
    \notag \\ 
\omega_{\mcf{J}}  &\leq \omega_{\mcf{H}} + O(  \sqrt{\rho_{\Delta}} ) 
    \notag \\ 
\varepsilon_{\mcf{J}}(\rho_{\mathrm{Dec}} ) &\leq  \varepsilon_{\mcf{H}}(\rho_{\mathrm{Dec}}+ \rho_{\Delta})  + e^{-O(n \delta^2 )}
    \notag \\
\alpha^*_{\mcf{J}}(\rho_{\mathrm{Dec}}, \rho_{\mathrm{Adv}}) 
    &\leq 
 e^{ -O(n \rho_{\Delta}^2)}.
    \notag
\end{align}
\end{cor}

\begin{remark}
Corollary~\ref{cor:1stcode} is a corollary of Theorem~\ref{thm:1stcode} located in Appendix~\ref{app:gen}.
Theorem~\ref{thm:1stcode}, unlike the above corollary, does not fix the injection noise power, tolerance, or the values in $\mcf{K}.$ 
Furthermore, Theorem~\ref{thm:1stcode} specifies the error terms instead of using order terms.  
\end{remark}

\begin{proofsketch}
The proof of Theorem~\ref{thm:1stcode} is found in Appendix~\ref{app:1stcode}, and Corollary~\ref{cor:1stcode} trivially follows.

Proving the rate is immediate, since it is unchanged from the original code. 

For the average power, we have to deal with the deterministic value of $\mbf{t}(M)$ added to the code, in particular analyzing the probability that a spurious value of $\mbf{t}(m)$ is chosen with a large amount of correlation with the related $\mbf{x}(m).$

For the probability of error, we have to consider both the probability of error of the original decoder with the added noise and $\mbf{t}$ as well as the probability of error introduced with the detector. 
To upper bound the probability of error of the original decoder, we use the fact that the randomly chosen value of $\mbf{t}$ plus the message-dependent additive white Gaussian noise terms is effectively a message-independent additive white Gaussian noise term with variance $\rho_{\Delta}.$
Hence, the error averaged over all possible choices of $\mbf{t}(M)$ is $\varepsilon_{\mcf{H}}(\rho_{\mathrm{Dec}} + \rho_{\Delta})$.
Using Hoeffding's inequality, it follows that the random choice of $\mbf{t}$ must yield a probability of error close to the average. 
On the other hand, the probability of error of the detector is straightforward to calculate since, under no manipulation, the detector is checking to see if a sum of independent random variables has the correct mean. 

Finally for the probability of targeted false authentication, we note that if the adversary does try to attack, then the distribution of the received sequence at the decoder will consist of independent Gaussian random variables where the variance of the $i$th coordinate is
$$ \tau_i(m) = \tau^\star (f_i(m)) := \frac{f_i^2(m) \rho_{\Delta} \rho_{\mathrm{Adv}}}{f_i^2(m) \rho_{\Delta} + \rho_{\mathrm{Adv}}} + \rho_{\mathrm{Dec}} ,$$
and the mean is of the adversary's choosing. 
For visualization purposes, note that when $\rho_{\Delta}$ becomes small this variance term converges to $f_i^2(m) \rho_{\Delta} + \rho_{\mathrm{Dec}}.$ 
By properties of the overlay code though, for each message and alternative message, there exists one set of overlay output coordinates whose output for the decoded message is less than or equal to an alternative message. 
The probability of detecting this increase in noise variance (under the assumption that the decoded message is not the one transmitted by the encoder is calculated and used to determine the probability of detecting the adversary's manipulation. 

\end{proofsketch}

While Code Modification~\ref{code:addnoise} does not allow the adversary to impersonate any specific message, it does not guarantee that the adversary cannot impersonate any message at all. 
This difference is made plain by referring to the operational definitions and observing again that
\begin{align}
 &\Pr \left( \hat m(\mbf{X}(a) + \mbf{G}_{\mathrm{Dec}} + \mbf{Z}(\mbf{V},a) ) \notin \{ a,\mbf{!}\}   \right)  \notag \\
 &\quad =  \sum_{b \in \mcf{M} \setminus \{a , \mbf{!}\}}  \Pr \left( \hat m(\mbf{X}(a) + \mbf{G}_{\mathrm{Dec}} + \mbf{Z}(\mbf{V},a) ) = b    \right) . \label{eq:ai0}
\end{align}
While Code Modification~\ref{code:addnoise} produces codes such that each summand $ \Pr \left( \hat m(\mbf{X}(a) + \mbf{G}_{\mathrm{Dec}} + \mbf{Z}(\mbf{V},a) ) = b    \right)$ is small, it does not guarantee the production of a code for which the sum itself is small.

Some reflection, though, shows that the case where the summand is small but this sum is not can only occur if there is (in some sense) a densely packed set of decoding regions. 
Under this notion, it makes sense to randomly decimate the message set, similar to how (and why) Ahlswede and Dueck~\cite{bcgc} chose to demonstrate the local strong converse. 
While this does reduce the rate of the code, only a negligible amount of loss (in terms of rate) is needed to guarantee the decoding regions are much less dense. 

We will resume with a slightly more formal description of why this works after we introduce the coding modification.
For now, we must mention that the amount of decimation the message set needs is dependent on operational measures of the underlying code. 
Therefore, to improve readability we have opted to produce a simplified version of the code modification here, and leave the more precise result for Appendix~\ref{app:gen}.

\begin{codingmodcor}[Code Modification~\ref{code:decimate}]\label{code:decimate_alt}
Suppose 
\begin{itemize} 
\item deterministic code $\mcf{H} = (\mbf{x}: \mcf{M} \rightarrow \mbcf{R}$, $\hat m : \mbcf{R} \rightarrow \mcf{M})$, 
\item injection noise power $\rho_{\Delta}= o(1)$, 
\item tolerance $\delta = o(\rho_{\Delta})$, 
\item $\left( \frac{1}{n} \log |\mcf{M}|, \mcf{K}, \gamma \right)$-overlay code $\mbf{f}:\mcf{M} \rightarrow \mbcf{\tilde K}$, for finite $\mcf{K} \subset [0,1)$ and $\gamma \in \left(\frac{1}{2},1\right)$,
\end{itemize} 
are given.

First apply Code Modification~\ref{code:addnoise} to code $\mcf{H}$, to obtain code $ \mbf{X}': \mcf{M} \rightarrow \mbcf{R}$, $\hat m' : \mbcf{R} \rightarrow \mcf{M} \cup \{\mbf{!}\}$. 
Next, select $\mcf{M}^\ddagger$ uniformly at random from $\left( \begin{matrix} \mcf{M} \\ \left \lfloor \exp (nr^\ddagger) \right \rfloor  \end{matrix} \right)$, where 
$$r^\ddagger = r_{\mcf{H}} - O\left(\rho_{\Delta}^2+\frac{\log n}{n} \right) .$$
Define the modified encoder $\mbf{X}^\ddagger : \mcf{M}^\ddagger \rightarrow \mbcf{R}$ by
$$\mbf{X}^\ddagger (M) = \mbf{X}'(M) .$$
Define the modified decoder $\hat m^\ddagger : \mbcf{R} \rightarrow \mcf{M}^\ddagger \cup \{\mbf{!}\}$ by
$$\hat m^\ddagger (\mbf{Y}) = \begin{cases} \hat m'(\mbf{Y}) & \text{ if } \hat m'(\mbf{Y}) \in \mcf{M}^\ddagger  \\
\mbf{!} & \text{ else}
\end{cases}.$$
The resulting modified code is given by $\mbf{X}^\ddagger, \hat m^\ddagger.$
\end{codingmodcor}

\begin{remark}
Code Modification Corollary~\ref{code:decimate_alt} is a corollary of Code Modification~\ref{code:decimate} located in Appendix~\ref{app:gen}.
There, the decimation terms are made explicit. 
\end{remark}

\begin{remarkstar}
Decimating the message set reduces the rate of the code. 
\end{remarkstar}

We now return to a more formal description of why this works, which follows from two important facts.
First, decimating the message set will not impact the maximum probability of targeted false authentication for any two non-decimated messages.
Second, by decimating the message set to $\mcf{M}^\ddagger$, the probability of false authentication for a given encoded message $a \in \mcf{M}$ and fixed adversary function $\mbf{Z}$ can be written as
\begin{equation}\label{eq:adint}
\sum_{b \in \mcf{M} \setminus \{a , \mbf{!}\}}  \idc{b}{\mcf{ M}^{\ddagger}}\Pr \left( \hat m(\mbf{X}(a) + \mbf{G}_{\mathrm{Dec}} + \mbf{Z}(\mbf{V},a) ) = b    \right).
\end{equation}
Equation~\eqref{eq:adint}, when considered jointly with the decimated message set $\mcf{M}^\ddagger$ being randomly chosen, takes a form whose concentration is analytically tractable.
More specifically Equation~\eqref{eq:adint} should with high probability be close to the mean, which is at most $|\mcf{M}^{\ddagger}|/|\mcf{M}|$ since this is the probability a message is not decimated.

The above intuition is overly-simplistic because all possible attacks must be simultaneously considered.
Nevertheless, the technique is sufficient to prove the next theorem/corollary. 

\begin{cor}[Theorem~\ref{thm:2ndcode}]\label{cor:2ndcode}
Setting injection noise power $\rho_{\Delta} = o(1) $ and tolerance $\delta = o(\rho_{\Delta})$,  then for all 
\begin{itemize} 
\item deterministic codes $\mcf{H} = (\mbf{x}: \mcf{M} \rightarrow \mbcf{R}$, $\hat m : \mbcf{R} \rightarrow \mcf{M})$ with rate $r_{\mcf{H}} = \Omega( n^{-1} \log n) $, 
\item $\left( r_{\mcf{H}}, \mcf{K}, \gamma \right)$-uniform overlay code $\mbf{f}: \mcf{M} \rightarrow \mbcf{\tilde K}$ for any $\gamma \in (1/2, 1)$,
\item and large enough $n$ 
\end{itemize}
Code Modification~\ref{code:decimate} with high probability yields a code $\mcf{J} = (\mbf{X}^\ddagger: \mcf{M}^\ddagger \rightarrow \mbcf{R}$, $\hat m^\ddagger : \mbcf{R} \rightarrow \mcf{M}^\ddagger \cup \{\mbf{!}\})$ such that 
\begin{align}
r_{\mcf{J}} &\geq r_{\mcf{H}} - O\left(\rho_{\Delta}^2+\frac{\log n}{n} \right)
    \notag \\ 
\omega_{\mcf{J}}  &\leq \omega_{\mcf{H}} + O( \sqrt{\rho_{\Delta}} ) 
    \notag \\ 
\varepsilon_{\mcf{J}}(\rho_{\mathrm{Dec}} ) &\leq  \varepsilon_{\mcf{H}}(\rho_{\mathrm{Dec}}+ \rho_{\Delta})  + e^{-O(n \delta^2)}
    \notag \\ 
\alpha_{\mcf{J}}(\rho_{\mathrm{Dec}},\rho_{\mathrm{Adv}} ) &\leq  e^{-O(n \rho_{\Delta}^2)}.
    \notag
    \notag
\end{align}
\end{cor}

\begin{remark}
Corollary~\ref{cor:2ndcode} is a corollary of Theorem~\ref{thm:2ndcode} located in Appendix~\ref{app:gen}.
Theorem~\ref{thm:2ndcode}, unlike the above corollary, does not fix the injection noise power, tolerance, or the values in $\mcf{K}.$ 
Furthermore, Theorem~\ref{thm:2ndcode} specifies the error terms instead of using order terms.
\end{remark}

\begin{proofsketch}
The proof of Theorem~\ref{thm:2ndcode} is found in Appendix~\ref{app:2ndcode}; note that it relies on elements of the proof of Theorem~\ref{thm:1stcode} since Code Modification~\ref{code:decimate} relies on Code Modification~\ref{code:addnoise}.

The rate and power for the new code are straightforward, while the probability of error calculation essentially follows from Hoeffding's inequality.

The major difficulty in the proof is proving the bound on the probability of false authentication. 
As the first step in proving this bound, we recall a result from the proof of Theorem~\ref{thm:1stcode}; specifically, that the decoder's observation when conditioned on a particular message, adversary observation, and adversary attack is equal to a sequence of independent random variables with the mean of the adversary's choosing but the variance fixed, i.e.,
$$\mbf{Y}|\{ M,\mbf{Y} ,\mbf{Z} = m, \mbf{v},  \mbf{z} \} = \mbf{G}_{\mbf{\tau}(m)} + \mbf{u}(m,\mbf{v},\mbf{z}),$$
where $\mbf{u}(m,\mbf{v},\mbf{z})$ is an arbitrary function (whose specification is unimportant for this proof) and
$$ \tau_i(m) = \frac{f_i^2(m) \rho_{\Delta} \rho_{\mathrm{Adv}}}{f_i^2(m) \rho_{\Delta} + \rho_{\mathrm{Adv}}} + \rho_{\mathrm{Dec}} $$
for each symbol $i \in \{1,\dots,n\}.$
Clearly, we can effectively ignore the values of $\mbf{V}$ and $\mbf{Z}$ by jointly considering $ \mbf{G}_{\mbf{\tau}(m)} + \mbf{\mu}$ for all $m \in \mcf{M}$ and $\mbf{\mu} \in \mbcf{R}$.

Now for any given $\mbf{\mu}\in \mbcf{R}$ and $m \in \mcf{M}^\ddagger$, we start by noting the probability of false authentication can be written 
\begin{equation}\label{eq:ps:thm:dec}
\sum_{b \in \mcf{M} \setminus \{m , \mbf{!}\}}  \idc{b}{\mcf{ M}^{\ddagger}}\Pr \left( \hat m'( \mbf{G}_{\mbf{\tau}(m)} + \mbf{\mu})  = b    \right),
\end{equation}
where $\hat m'$ is the modified decoder resulting from the application of Code Modification~\ref{code:addnoise} in Code Modification~\ref{code:decimate}.
Using a modified version of the Hoeffding lemma we then bound the concentration of equation~\eqref{eq:ps:thm:dec}.
The problem that remains is to extend above concentration to simultaneously work for all $\mbf{\mu} \in \mbcf{R}$.

Here we take a divide-and-conquer approach by separately considering the sets of $\mbf{\mu} \in \mbcf{U}^\dagger$ and $\mbf{\mu} \notin \mbcf{U}^\dagger$, where $\mcf{U}^\dagger$ is a bounded interval on the real number line. 
These bounds are set sufficiently large so that $\mbf{\mu} \notin \mbcf{U}^\dagger$ guarantees that for the coordinate such that $\mu_i \notin \mcf{U}^\dagger$, the probability of passing the detector for each message is less than $e^{-n r_{\mcf{H}} - O(n \rho_{\Delta}^2)},$ and hence the probability of passing any message detector is less than $e^{-O(n\rho_{\Delta}^2)}.$
For $\mbf{\mu} \in \mbcf{U}^\dagger$, we show that there exists a finite set $\mbcf{U}^\ddagger \subset \mbcf{R}$ such that bounding all $\mbf{\mu} \in \mbcf{U}^\ddagger$ will suffice to bound all $\mbf{\mu} \in \mbcf{U}^\dagger$. 
From there, we use the union bound to simultaneously guarantee the concentration of all $\mbf{\mu} \in \mbcf{U}^\ddagger$ (hence all $\mbf{\mu} \in \mbcf{U}^\dagger$) and all $m \in \mcf{M}.$

\end{proofsketch}


At this point, it is important to reflect on the form of Theorem~\ref{thm:2ndcode}/Corollary~\ref{cor:2ndcode}.
Specifically, consider Corollary~\ref{cor:2ndcode} where $\delta$ is chosen such that $\lim n\delta^2 = \infty$. 
For example $\rho_{\Delta} = \sqrt[-4]{n} \log n$ and $\delta = \sqrt[-4]{n}.$
In this case, the code modifications have necessitated a loss in rate, an increase in power, and require the code to be operational at a larger noise level than the original code. 
However, each of these changes disappear as $n$ increases, and hence the rate converges back to the original rate, the new power converges to the original power, and the level of noise the code must be robust against converges to the original noise level.
Suppose then we start with a capacity-achieving sequence of codes with average power $\omega - O(\sqrt{\rho_{\Delta}}),$ and which are robust to a noise variance of $\rho_{\Delta} + \rho_{\mathrm{Dec}}$. 
Applying Theorem~\ref{thm:2ndcode}/Corollary~\ref{cor:2ndcode} should give us a sequence of codes with rate 
$$\lim_{n \rightarrow \infty} \frac{1}{2} \log \left( 1 + \frac{\omega - O(\sqrt{\rho_{\Delta}})}{\rho_{\Delta} + \rho_{\mathrm{Dec}}} \right) - o(1) = \frac{1}{2} \log \left( 1 + \frac{\omega}{\rho_{\mathrm{Dec}}} \right), $$
which is capacity. 
At the same time, plugging the values into the maximum probability of false authentication yields
$$\lim_{n \rightarrow \infty} \alpha_{\mcf{J}}(\rho_{\mathrm{Dec}},\rho_{\mathrm{Adv}} ) \leq  \lim_{n \rightarrow \infty}  e^{-O(n \rho_{\Delta}^2)} = 0,$$
and thus we have the ability to authenticate.
This essentially proves the following theorem.

\begin{theorem}\label{thm:cap}
$$c(\rho,\rho_{\mathrm{Dec}}, \rho_{\mathrm{Adv}}) = \begin{cases} \frac{1}{2} \log \left( 1 + \frac{\rho}{\rho_{\mathrm{Dec}}} \right)  & \text{if } \rho_{\mathrm{Adv}}>0 \\ 0 & \text{else.} \end{cases}$$
\end{theorem}
\begin{proofsketch}
The proof Theorem~\ref{thm:cap} is found in Appendix~\ref{app:cap} and is essentially a more formal version of the discussion preceding the theorem. 
Additionally, we show that if $\rho_{\mathrm{Adv}} = 0$ then the capacity is zero. 
This is somewhat obvious since the adversary knows the encoder's output perfectly in this case.
\end{proofsketch}

Notice that the capacity experiences a sharp jump at $\rho_{\mathrm{Adv}} = 0,$ but is otherwise independent of the value.
From a practical perspective, this is ideal.
A perfect continuous channel is a physical impossibility, thus allowing us to assume that $\rho_{\mathrm{Adv}}$ is greater than zero. Hence, our result implies that in practical wireless scenarios, information-theoretic authentication is possible without use of a secret key.
It is also important to observe that the code modifications themselves do not rely on knowledge of the adversary's channel.

Interestingly, our results indicate that obtaining information-theoretic authentication from a channel differs significantly from  obtaining information-theoretic secrecy from a channel. 
Indeed, all practically relevant schemes for the wiretap channel, dating back to Wyner's seminal work~\cite{wyner75wtc}, require both knowledge of the adversary's channel as well certain guarantees on this channel which make implementation a difficult proposition.
In the relevant analog to our model\footnote{Specifically, from Figure~\ref{fig:chan} remove the message side information given to the adversary and remove the adversary's output.}, information-theoretic secrecy cannot be guaranteed when the noise to the adversary is less than the noise to the decoder. 
This is not an impediment to information-theoretic authentication though, as our results demonstrate; {importantly, knowledge of the message is distinct from knowledge of the transmitted sequence.}

In the next section we will discuss the path forward in more detail. 
Among other things, we will discuss unexplored alternatives for implementation, barriers to practical implementation, difficulties in other channels, and different implementation scenarios.

\section{Discussion \& Future Directions}\label{sec:futures}

While we derive a scheme that leads to information-theoretic authentication, there remains much to be done.
It is worth discussing these remaining questions with some candor, so that those so motivated have a clear understanding of areas for improvement. 
We also provide here further discussion on the distinction (beyond the obvious) between secret-key based authentication and what we accomplish here.

\subsection{Overlay code improvements}

When first formulating overlay codes, the goal was to ensure the unique relationship of the output symbols for different messages.
In the construction, there were a number of different parameters that could have been varied. In particular: the number of coordinates for a given output concentration, the overlap amount per coordinate, and the output levels themselves (i.e., $\mcf{K}$). 
To simplify our analysis, we chose to fix the first two considerations, while leaving $\mcf{K}$ variable.
Surprisingly, the actual values for $\mcf{K},$ while they do impact the efficiency of the authentication scheme, are actually rather immaterial to achieving authentication. 

Further analysis showed the optimal values of $\mcf{K}$ depend on the value of $\rho_{\mathrm{Adv}}.$ 
As a result, we chose not to optimize over $\mcf{K}$ since important to our claims is that the value of $\rho_{\mathrm{Adv}}$ need not be known when constructing the code.
During the review process, the question of the optimal value of $\mcf{K}$ was raised. 
To that end, when $\rho_{\Delta} = o(1)$ the optimal choice of $\mcf{K}= \{0,k_{(1)},\dots, k_{(|\mcf{K}|-1)} \}$ converges to 
$$ k_{(a)} = \left[\frac{(1-c\gamma)}{c(1-\gamma)}\right]^{a -1}   \frac{\rho_{\mathrm{Dec}}}{\rho_{\Delta}} ,$$
where
$$c = \frac{\sqrt[|\mcf{K}|]{\rho_{\mathrm{Dec}}} }{\gamma \sqrt[|\mcf{K}|]{\rho_{\mathrm{Dec}}}  + (1-\gamma) \sqrt[|\mcf{K}|]{ \rho_{\Delta}+\rho_{\mathrm{Dec}} }},$$
and yields a maximum probability of false authentication (subject to our analysis) of essentially
$$\exp\left(-\frac{1}{8}  (1-\gamma) [1 - (1+\delta) c ]^2 \right).$$
Still, use of this asymptotically optimal value did not simplify our analysis and hence was not instituted.

This does, however, raise the question of what is being lost (in terms of authentication ability) by choosing a sub-optimal values for $\mcf{K}$. Specifically, it would be interesting to quantify that loss in such a way as to allow for choosing $\mcf{K}$ to minimize the maximum of the maximum of the probability of false authentication. 
Additionally, it remains an open question whether allowing variable $\gamma$ and variable coordinates per output symbol could further improve the final results.

\subsection{Practical implementation of code modifications}\label{sec:fut:pi}

To enable authentication, message-dependent noise must be added and then certain distance properties between the codewords must be ensured.
These two tasks appear here as Code Modifications~\ref{code:addnoise} and~\ref{code:decimate}.
Our original intent was practicality in these code modifications; we were moderately successful with regards to Code Modification~\ref{code:addnoise}, but not so with~\ref{code:decimate}. 
That Code Modification~\ref{code:addnoise} could be reasonably implemented guided our decision to include here the non-asymptotic versions of Theorem~\ref{thm:1stcode} and~\ref{thm:2ndcode}. 
Still, it is worthwhile to discuss alternatives to our code modifications that could allow for an analytical bounds on the operational parameters, as well as a practical implementation.

For Code Modification~\ref{code:addnoise}, the only real concern in terms of practicality is the construction of the $\mbf{t}$ function.
Indeed, since the initial decoder is used in the first stage of the updated decoder, the output of the decoder can be used to determine what the appropriate value of $\mbf{t}$ should be for the estimated message. 
It is worth mentioning that we suspect that setting $\mbf{t}$ equal to zero will suffice in most cases. 
Our suspicion derives from the fact that $\mbf{t}$ is only needed to ensure that the code appears to have uniform noise across all coordinates. 
In practical decoders though, less noise per symbol is usually to the decoder's benefit.
Setting $\mbf{t}$ to zero would yield $\omega_{\mcf{J}} \leq \omega_{\mcf{H}} + \rho_{\Delta}$, with the rate and probability of targeted false authentication remaining as in Theorem~\ref{thm:1stcode}.
On the other hand, the average arithmetic error could be estimated empirically.
Hence, this should result in a practical implementation of Code Modification~\ref{code:addnoise} for which Theorem~\ref{thm:1stcode} is relevant.

Code Modification~\ref{code:decimate}, on the other hand, cannot be directly implemented as currently stated. 
Choosing such a large subset uniformly at random from the set of all such subsets is clearly impossible in practice.
There may of course be feasible alternatives. For instance, the subset selection could be accomplished using a universal hash function, and the Hoeffding concentration analysis replaced with one deriving from the leftover hash lemma.
Alternatively, it may be possible to show that some codes do not actually require a rate reduction.
Indeed, our analysis for Code Modification~\ref{code:decimate} relies heavily on the maximum probability of targeted false authentication established by Code Modification~\ref{code:addnoise}. 
But the adversary can only obtain this maximum by choosing a very specific output, and cannot obtain it for multiple alternative messages at one time.
As a result, it seems likely that a more sophisticated analysis, using the amount of perturbation from the optimal output, could yield a maximum distance between codewords required for there to be a successful attack. 
Ensuring that the code's minimum distance was greater than this maximum would be sufficient to skip Code Modification~\ref{code:decimate} entirely.


\subsection{Comparison with secret key-based authentication}

The major advantage our authentication scheme has over one that is secret key-dependent is that the secret key becomes a finite resource when the channel to the adversary is better than the channel to the decoder.
Hence, in some channel models, our scheme could operate in perpetuity while one which is key-based would have a finite life span. 
That this is particularly true in any case where the adversary has a better channel is shown in Graves et al.~\cite{graves2020secret} whose converse proves the key has a finite duration of use. 

On the other hand, secret key-based authentication still allows for two advantages over the non-secret key-based authentication of this paper. 
First, it is still operational when there is no noise over the channel to the adversary\footnote{A physical impossibility.} and when the adversary knows, and can therefore cancel, the decoder's noise\footnote{Also a physical impossibility.}.
Second, and more important, secret key-based authentication experiences a better trade-off between rate loss and how quickly the probability of false authentication converges to zero.

For secret key-based authentication, we know that there must exist a trade-off between the channel capacity and the exponent for the probability of false authentication due to the converse results from Graves and Wong~\cite{graves2019inducing} and Graves et al.~\cite{graves2020secret}.
For some measures of false authentication, this trade-off is linear, and in that sense the message rate and probability of false authentication must share the channel capacity. 

Our results do not allow for this type of trade off.
That is, while our results require a reduction in rate from the channel capacity in order to achieve authentication, the exponent for the probability of false authentication is at most $\exp(-o(n))$ whereas secret key-based authentication allows $\exp(-O(n)).$
If we assume that the encoder knows the channel to the adversary\footnote{This comparison to secret key-based authentication is not entirely fair, since knowledge of the channel to the adversary is not needed in that case.} then it is possible to also achieve $\exp(-O(n))$ with our results.
Indeed, this is because in this case we do not need $\rho_{\Delta}$ to vanish, but instead just be sufficiently small. 
Regardless, even under this unfair comparison, and further assuming the more generous result on the power constraint raised in Section~\ref{sec:fut:pi} and that the second code modification was unnecessary, to obtain a maximum probability of false authentication of $\exp(-O(n\rho_{\Delta}^2))$ requires that the difference between the maximum rate and capacity be at least
\begin{align}
&\frac{1}{2} \log \left( 1 + \frac{\rho}{\rho_{\mathrm{Dec}}} \right) - \frac{1}{2} \log \left( 1 + \frac{\rho- \rho_{\Delta}}{\rho_{\mathrm{Dec}} + \rho_{\Delta}} \right) 
    \notag \\ & \quad 
    =
\sum_{i = 1}^\infty \frac{1}{i} (c \rho_{\Delta})^i 
    \notag
\end{align}
where
$$c = \frac{\rho+\rho_{\mathrm{Dec}}}{\rho + \rho_{\mathrm{Dec}} +  \rho_{\Delta} \left( 1 + \frac{\rho}{\rho_{\mathrm{Dec}}} \right)}.$$
Hence, a loss of rate does not lead to a linear increase in the exponent of maximum probability of false authentication using our scheme.

\subsection{Higher order wireless channel approximations}
While Gaussian channels are great approximations for free-space fixed point single antenna communications\footnote{This point is discussed by Massey~\cite{massey1992deep} regarding deep-space communications.}, there exist other scenarios of wireless communications with their own corresponding best channel approximations. 
Some of these alternative channels consider \emph{multi-input multi-output} (MIMO) antenna arrays, fading channels, and multi-path channels. 

Outright, we do not see any reason that the overlay code concept cannot be modified and applied to these channels to create codes that provide information-theoretic authentication. 
However, any such modification will be highly dependent on the assumptions placed on the encoder and decoder with regards to knowledge of their own channels.

\section{Conclusion}\label{sec:con}

In this work we have shown that physical layer authentication is possible for a channel that models wireless communication.
Not only is physical layer authentication possible, but our scheme can be used to detect any adversary as long as the block length is sufficiently large and the adversary does not have access to a completely noiseless copy of the transmission.
Our scheme achieves this by adding artificial noise into the system using the novel concept of overlay codes.
This approach allows for authentication by forcing the adversary to remove the added noise when they hope to insert a fake message of their own.

Although random coding elements were used in the proofs, many of the difficulties in practical implementation do not exist in our modular scheme.
That is, only the encoder needs to be constructed, since part of the concept of the modular scheme is that the message can still be decoded using the original decoder {(see Section \ref{sec:fut:pi})}. 
Furthermore, our modular scheme establishes that every deterministic channel code has a variant which can provide physical layer authentication.
We expect this to lower the implementation barrier since we therefore do not require a completely new channel code be added to the system design.

{Open problems include those outlined in Section \ref{sec:futures}, as well as investigating further scenarios where adding artificial noise can provide authentication.}


\bibliographystyle{IEEEtran}
\bibliography{this,this2}

    \appendices

\section{Non-asymptotic versions of Corollaries~\ref{cor:1stcode} and~\ref{cor:2ndcode} and Code Modification Corollary~\ref{code:decimate_alt}}\label{app:gen}
\begin{theorem} \label{thm:1stcode}
For all 
\begin{itemize} 
\item deterministic codes $\mcf{H} = (\mbf{x}: \mcf{M} \rightarrow \mbcf{R}$, $\hat m : \mbcf{R} \rightarrow \mcf{M})$, 
\item injection noise power $\rho_{\Delta} \in (0,\infty)$, 
\item tolerance $\delta \in (0,1)$, 
\item and  $\left( \frac{1}{n} \log |\mcf{M}|, \mcf{K}, \gamma \right)$-overlay codes $\mbf{f}: \mcf{M} \rightarrow \mbcf{\tilde K}$, for finite $\mcf{K} \subset [0,1)$ and $\gamma \in \left(\frac{1}{2},1\right)$,
\end{itemize}
Code Modification~\ref{code:addnoise} with high probability yields a code $\mcf{J} = (\mbf{X}' : \mcf{M} \rightarrow \mbcf{R}$, $\hat m': \mbcf{R} \rightarrow \mcf{M} \cup \{\mbf{!} \})$ such that   
\begin{align}
r_{\mcf{J}} &= r_{\mcf{H}} 
    \notag \\ 
\omega_{\mcf{J}}  &\leq \omega_{\mcf{H}} + 2\sqrt{ 2 \omega_{\mcf{H}} \rho_{\Delta} (r_{\mcf{H}} + 1) } 
    \notag \\ & \quad 
+ \!\rho_{\Delta} \! \left(\! 1 \!+ \!8 |\mcf{\tilde K}| \left[ r_{\mcf{H}} \!+\!1\! + \!\frac{\log |\mcf{ K}|}{n} \right] \right)
    \notag \\ 
\varepsilon_{\mcf{J}}(\rho_{\mathrm{Dec}} ) &\leq  \varepsilon_{\mcf{H}}(\rho_{\mathrm{Dec}}+ \rho_{\Delta})  + \sqrt{\frac{n}{2} e^{-nr_{\mcf{H}}}}
+ |\mcf{K}| e^{-\frac{1}{8} \ell \delta^2} 
    \notag \\
\alpha^*_{\mcf{J}}(\rho_{\mathrm{Dec}}, \rho_{\mathrm{Adv}}) 
    &\leq 
 e^{ -\frac{1}{8}  \ell (1-\gamma)  \lambda^2 } + e^{-\frac{1}{8} \ell\gamma \lambda^2}
    \notag
\end{align}
where
\begin{align}
\lambda &= \max\left( 0 ,  \min_{k \in \mcf{K}} 1 - \frac{(1+\delta)(k^2\rho_{\Delta} + \rho_{\mathrm{Dec}})}{\gamma \tau^\star(k) + (1-\gamma) \tau^\star(d_k) } \right)
    \notag \\
\tau^\star(a) &= \frac{a^2 \rho_{\Delta} \rho_{\mathrm{Adv}}}{a^2 \rho_{\Delta} + \rho_{\mathrm{Adv}}} + \rho_{\mathrm{Dec}} 
    \notag \\ 
d_k &= \min\{ d \in \mcf{\tilde K} | d > k \}.\notag
\end{align}
\end{theorem}

\begin{codingmod}\label{code:decimate}
Suppose 
\begin{itemize} 
\item deterministic code $\mcf{H} = (\mbf{x}: \mcf{M} \rightarrow \mbcf{R}$, $\hat m : \mbcf{R} \rightarrow \mcf{M})$, 
\item injection noise power $\rho_{\Delta}\in (0,\infty)$, 
\item tolerance $\delta\in(0,1)$, 
\item $\left( \frac{1}{n} \log |\mcf{M}|, \mcf{K}, \gamma \right)$-overlay code $\mbf{f}:\mcf{M} \rightarrow \mbcf{\tilde K}$, for finite $\mcf{K} \subset [0,1)$ and $\gamma \in \left(\frac{1}{2},1\right)$,
\end{itemize} 
are given.

First,  apply Code Modification~\ref{code:addnoise} to code $\mcf{H}$ and let $ \mbf{X}': \mcf{M} \rightarrow \mbcf{R}$, $\hat m' : \mbcf{R} \rightarrow \mcf{M} \cup \{\mbf{!}\}$ be the result. 
Next, select $\mcf{M}^\ddagger$ uniformly at random from $\left( \begin{matrix} \mcf{M} \\ \lfloor \exp(n r^\ddagger) \rfloor \end{matrix} \right)$, where $r^\ddagger$ 
$$r^\ddagger = (1-n^{-1}) r_{\mcf{H}} -  \frac{(1-\gamma)\ell}{4n}\lambda^2 - \frac{2 +  \log 2 \theta}{n} $$
and $\lambda,~\tau^\star,$ and $d_k$ are as defined in Theorem~\ref{thm:1stcode}, while 
\begin{align}
\theta & = \max \left( \!\!1,\!\! \sqrt{3n \left[ \omega_{\mcf{K}}  + (\rho_{\Delta} + \rho_{\mathrm{Dec}}) \left(1 + \delta + 2 \lambda^2 + 2 r_{\mcf{J}} \right)\right] } \right)
    \notag.
\end{align}
Define the modified encoder $\mbf{X}^\ddagger : \mcf{M}^\ddagger \rightarrow \mbcf{R}$ by
$$\mbf{X}^\ddagger (M) = \mbf{X}'(M) .$$ 
Define the modified decoder $\hat m^\ddagger : \mbcf{R} \rightarrow \mcf{M}^\ddagger \cup \{\mbf{!}\}$ by
$$\hat m^\ddagger (\mbf{Y}) = \begin{cases} \hat m'(\mbf{Y}) & \text{ if } \hat m'(\mbf{Y}) \in \mcf{M}^\ddagger  \\
\mbf{!} & \text{ else}
\end{cases}.$$

The new modified code is $\mbf{X}^\ddagger, \hat m^\ddagger.$
\end{codingmod}

\begin{theorem}\label{thm:2ndcode}
For all 
\begin{itemize} 
\item deterministic codes $\mcf{H} = (\mbf{x}: \mcf{M} \rightarrow \mbcf{R}$, $\hat m : \mbcf{R} \rightarrow \mcf{M})$, with rate
$$ r_{\mcf{H}}   \geq \frac{(1-\gamma)\ell}{4(n-1)}\lambda^2 + \frac{2 +  \log  4n\theta}{n-1}  $$
\item injection noise power $\rho_{\Delta}\in (0,\infty)$, 
\item tolerance $\delta\in(0,1)$, 
\item and $\left( \frac{1}{n} \log |\mcf{M}|, \mcf{K}, \gamma \right)$-overlay codes $\mbf{f}:\mcf{M} \rightarrow \mbcf{\tilde K}$, for finite $\mcf{K} \subset [0,1)$ and $\gamma \in \left(\frac{1}{2},1\right)$,
\end{itemize}
with high probability Code Modification~\ref{code:decimate} yields a code $\mcf{J} = (\mbf{X}^\ddagger: \mcf{M}^\ddagger \rightarrow \mbcf{R}$, $\hat m^\ddagger : \mbcf{R} \rightarrow \mcf{M}^\ddagger \cup \mbf{!})$ such that 
\begin{align}
r_{\mcf{J}} &\geq   r_{\mcf{H}} -  \frac{(1-\gamma)\ell}{4n}\lambda^2 - \frac{r_{\mcf{H}} + 2 +  \log 4n \theta}{n} 
    \notag \\ 
\omega_{\mcf{J}}  &\leq \omega_{\mcf{H}} + 2\sqrt{ 2 \omega_{\mcf{H}} \rho_{\Delta} (r_{\mcf{H}} + 1) } 
    \notag \\ & \quad 
+ \!\rho_{\Delta} \! \left(\! 1 \!+ \!8 |\mcf{\tilde K}| \left[ r_{\mcf{H}} \!+\!1\! + \!\frac{\log |\mcf{ K}|}{n} \right] \right)
    \notag \\ 
\varepsilon_{\mcf{J}}(\rho_{\mathrm{Dec}} ) &\leq  \varepsilon_{\mcf{H}}(\rho_{\mathrm{Dec}}+ \rho_{\Delta})  + \sqrt{2ne^{-nr_{\mcf{H}}} }
+ |\mcf{K}| e^{-\frac{1}{8} \ell \delta^2} 
    \notag \\
\alpha_{\mcf{J}}(\rho_{\mathrm{Dec}},\rho_{\mathrm{Adv}} ) &\leq  \left(2n + \frac{1}{2\sqrt{ n \rho_{\mathrm{Dec}}}}  \right)  e^{-\frac{1-\gamma}{8} \ell \lambda^2},
    \notag
    \notag
\end{align}
where 
$\theta$ is defined in Code Modification~\ref{code:decimate} and $\lambda$ is defined in Theorem~\ref{thm:1stcode}.

\end{theorem}

\section{Theorem~\ref{thm:codevid19} and Corollary~\ref{cor:wedontwantthemtogetofftheshipbecausethatmaydoubleournumbers}}\label{app:codevid19}

The proof of theorem and corollary rely on the following random code construction. 
\begin{coding}\label{cc:overlay}


Suppose finite set $\mcf{K}\subset [0,1)$ and positive number $\gamma \in ( 1/2, 1)$ are given. 

For convenience, for each $k\in \mcf{K}$ set
\begin{align} 
n_k &= n - \ell |\{j\in \mcf{K}| j<k \}|   , 
    \notag \\ 
\mcf{N}_k &= \{1,\dots,n_k\}, 
    \notag \\
\mcf{S}_k &= \left( \begin{matrix} \mcf{N}_k \\ \ell \end{matrix} \right) 
    \notag \\
r_k &= \frac{  \log \left \lfloor \exp\left(n_k \left| \mathbb{I}_2\left(\gamma  \middle|\middle| \frac{\ell}{n_k} \right) - \frac{1}{3n_k} - \frac{2}{n_k}\log  n_k \sqrt{\ell} \right|^+\right)\right\rfloor}{n_k}
    \notag.
\end{align}
For $k = 1$, let $n_k$ and $\mcf{N}_k$ be defined as above, but let $\mcf{S}_k = \left( \begin{matrix} \mcf{N}_k \\ n_k \end{matrix} \right)$ and $r_k = 0.$

Next for each $k \in \mcf{ K} $ let $\mcf{M}_k = \{1,\dots, e^{n_{k}r_{k}}\}$, allowing that $\mcf{M} = \times_{k\in \mcf{K}} \mcf{M}_k.$

Independently for each $k\in \mcf{K}$ and each $m_k \in \mcf{M}_k$, choose a set $\mcf{S}_k(m_k)$ uniformly at random from $\mcf{S}_k.$
Then, for each $m = \times_{k \in \mcf{K}} m_k \in \mcf{M}$ and $j \in  \{1,\dots, n\}$ set 
\begin{align}
&f_j(m) = k 
    \quad \Leftrightarrow \quad
j \in  g_{\mcf{N}_k \rightarrow \mcf{I}\left( \times_{ j \in \mcf{K} | j < k} m_j \right) }(\mcf{S}_k(m_k)) ,
    \notag
\end{align}
where $\mcf{I}\left( \times_{ j \in \mcf{K} | j < k} m_j \right)$ is defined recursively by
\begin{align}
    &
\mcf{I}\left( \times_{ j \in \mcf{K} | j \leq k} m_j \right)
    \notag \\ & \quad = 
\mcf{I}\left( \times_{ j \in \mcf{K} | j < k} m_j \right) - g_{\mcf{N}_k \rightarrow \mcf{I}\left( \times_{ j \in \mcf{K} | j < k} m_j \right) }(\mcf{S}_k(m_k)) 
    \notag 
\end{align}
with $\mcf{I}(\emptyset) = \{1,\dots,n\} $, and where $g_{\mcf{A}\rightarrow \mcf{B}}: \mcf{A} \rightarrow \mcf{B}$ is the lexicographical order-preserving mapping between two equal size sets of natural numbers.

\end{coding}

Prior to using this code construction to prove the theorem and corollary, we will present an example to make the construction more clear, as well as present a technical lemma in order to streamline the proof.

\subsection{Example overlay code construction}\label{app:ex}

Suppose $n=9$ and $\mcf{K} = \left\{ 0, 1/2 \right\}$, (hence $\ell = \left \lfloor \frac{9}{3} \right \rfloor = 3$) are given.
For simplicity, let rates $r_0,~r_{\nicefrac{1}{2}}$ and $r_1$ be such that $e^{9 r_0} = 4$, $e^{6 r_{\nicefrac{1}{2}}} = 3$, and $e^{3 r_1} = 1,$ yielding a total of $4\cdot 3 \cdot 1 = 12$ different messages, or a rate of $\frac{1}{9} \log 12.$
Note, that we do not need to specify $\gamma$ in this case since its only involvement in the code construction is choosing values for the rates.

Suppose the randomly selected subsets, $\mcf{S}_0(i)  \subset \{1,\dots, 9\}$ for $i \in \mcf{M}_0$ and $\mcf{S}_{\nicefrac{1}{2}}(j)\subset \{1,\dots, 6\}$ for $j \in \mcf{M}_{\nicefrac{1}{2}}$, are
\begin{align}
\begin{array}{rl} \mcf{S}_0(1) &= \{2, 7 , 8\} \\ \mcf{S}_0(2) &= \{1, 2 , 6\} \\ \mcf{S}_0(3) &= \{2,6,9\} \\ \mcf{S}_0(4) &= \{1,5,9\}\end{array}
\quad \text{ and } \quad 
\begin{array}{rl} \mcf{S}_{1/2}(1) &= \{2, 4 , 5\} \\ \mcf{S}_{1/2}(2) &= \{3, 4 , 6\} \\ \mcf{S}_{1/2}(3) &= \{1,3,5\} \end{array}
\notag
\end{align}
then the resulting code constructed is
\begin{align}
\begin{array}{c r c c c c c c c c c}
\mbf{f}(11) &= &1 &0 &\nicefrac{1}{2} &1 &\nicefrac{1}{2} &\nicefrac{1}{2} &0 &0 &1 
  \\
\mbf{f}(12) &= &1 &0 &1 &\nicefrac{1}{2} &\nicefrac{1}{2} &1 &0 &0 &\nicefrac{1}{2}
 \\
\mbf{f}(13) &= &\nicefrac{1}{2} &0 &1 &\nicefrac{1}{2} &1 &\nicefrac{1}{2} &0 &0 &1
     \\
\mbf{f}(21) &= &0 &0 &1 &\nicefrac{1}{2} &1 &0 &\nicefrac{1}{2} &\nicefrac{1}{2} &1 
     \\
\mbf{f}(22) &= &0 &0 &1 &1 &\nicefrac{1}{2} &0 &\nicefrac{1}{2} &1 &\nicefrac{1}{2} 
     \\
\mbf{f}(23) &= &0 &0 &\nicefrac{1}{2} &1 &\nicefrac{1}{2} &0 &1 &\nicefrac{1}{2} &1 
     \\
\mbf{f}(31) &= &1 &0 &\nicefrac{1}{2} &1 &\nicefrac{1}{2} &0 &\nicefrac{1}{2} &1 &0 
     \\
\mbf{f}(32) &= &1 &0 &1 &\nicefrac{1}{2} &\nicefrac{1}{2} &0 &1 &\nicefrac{1}{2} &0
     \\
\mbf{f}(33) &= &\nicefrac{1}{2} &0 &1 &\nicefrac{1}{2} &1 &0 &\nicefrac{1}{2} &1 &0
     \\
\mbf{f}(41) &= &0 &1 &\nicefrac{1}{2} &1 &0 &\nicefrac{1}{2} &\nicefrac{1}{2} &1 &0 
     \\
\mbf{f}(42) &= &0 &1 &1 &\nicefrac{1}{2} &0 &\nicefrac{1}{2} &1 &\nicefrac{1}{2} &0 
     \\
\mbf{f}(43) &= &0 &\nicefrac{1}{2} &1 &\nicefrac{1}{2} &0 &1 &\nicefrac{1}{2} &1 &0 
\end{array} \notag.
\end{align}
In more detail, take for example $\mbf{f}(32)$ which corresponds to $\mcf{S}_0(3) = \{2,6,9\}$ and $\mcf{S}_{\nicefrac{1}{2}}(2) = \{3,4,6\}$.
Here $\mbf{f}(32)$ is constructed by first assigning a $0$ to all indices in $\mcf{S}_0(3)$, after which these indices are removed from the pool of possible indices $\{1,2,3,4,5,6,7,8,9\}$ leaving indices $\{1,3,4,5,7,8\}$. 
Now considering the remaining indices ($\{1,3,4,5,7,8\}$) as an ordered set, of these the $\mcf{S}_{\nicefrac{1}{2}}(2)$ indices (the $\{3,4,6\}$-th smallest, i.e., $\{4,5,8\}$) are assigned a value of $\frac{1}{2}$, and all remaining unassigned indices ($\{1,3,7\}$) are given $1$.

Also from this example, the important aspect of the overlay code can be observed.
Namely, for any fixed message one of the sets of coordinates for the message which produce the same output (e.g., for $33$, $\{2,6,9\}$ produce $0$, $\{1,4,7\}$ produce $\nicefrac{1}{2}$, and $\{3,5,8\}$ produce $1$) is strictly not greater than the corresponding outputs produced for any alternative message.

\subsection{Technical lemma}

\begin{lemma}\label{lem:tec_codevid19}
For integers $a,b,c$, such that $a> b > c\geq \max(b -(a-b),1)$,
$$
-\log \frac{\left(\begin{matrix} b \\ c \end{matrix} \right)\left(\begin{matrix} a - b \\ b -c  \end{matrix} \right) }{\left(\begin{matrix} a \\ b \end{matrix} \right)} \geq a \mathbb{I}_2\left( \frac{c}{b} \middle|\middle| \frac{b}{a} \right) - \frac{1}{3} - 2 \log a .
$$
\end{lemma}
\begin{IEEEproof}
This lemma follows nearly directly from Robbins' remark\footnote{For all positive integers $n$,  $n! = \sqrt{2 \pi n} \left( \frac{n}{e} \right)^n e^{\zeta}$ for some $\zeta$ such that $\frac{1}{12n+1} \leq \zeta \leq \frac{1}{12n}$.} on Stirling's formula~\cite{robbins1955remark}, along with some basic algebra. 
More specifically 
\begin{align}
&-\log \frac{\left(\begin{matrix} b \\ c \end{matrix} \right)\left(\begin{matrix} a - b \\ b -c  \end{matrix} \right) }{\left(\begin{matrix} a \\ b \end{matrix} \right)}
    \notag \\ & \quad = 
c \log \frac{c}{b} + (b-c) \log \frac{b-c}{b} 
    \notag \\ & \quad \quad +
(a-2b+c) \log \frac{a-2b+c}{a-b} + (b-c) \log \frac{b-c}{a-b} 
    \notag \\ & \quad \quad - 
b \log \frac{b}{a} - (a-b) \log \frac{a-b}{a}  
    \notag \\ & \quad \quad - 
 \zeta' - \log \frac{b(a-b)}{(b-c)\sqrt{2 \pi ac (a-2b+c)}},
\end{align}
for some $\zeta'$ such that 
$$\zeta' \leq \frac{1}{12b} + \frac{1}{12(a-b)} + \frac{1}{12(a-b)+1} + \frac{1}{12b+1}, $$
by Robbins' remark.
Clearly though $\zeta' \leq \frac{1}{3}$ since $a>b\geq1$, while
$$\log \frac{b(a-b)}{(b-c)\sqrt{2 \pi ac (a-2b+c)}} \leq 2 \log a$$
due to the constraints placed on $a,b,c$ in the lemma statement.
To simplify the remainder of the statement recognize that 
\begin{align}
b \log \frac{b}{a} &= (b-c) \log \frac{b}{a} + c \log\frac{b}{a} \\
(a-b) \log \frac{a-b}{a}& = (a-2b+c) \log \frac{a-b}{a} 
   \notag \\ &\quad + 
(b-c) \log \frac{a-b}{a},
\end{align}
hence
\begin{align}
&c \log \frac{c}{b} + (b-c) \log \frac{b-c}{b} 
    \notag \\ &   +
(a-2b+c) \log \frac{a-2b+c}{a-b} + (b-c) \log \frac{b-c}{a-b} 
    \notag \\ &   - 
b \log \frac{b}{a} - (a-b) \log \frac{a-b}{a}  
    \notag \\ &  \quad = 
b\left[ \frac{c}{b} \log \frac{\frac{c}{b}}{\frac{b}{a}}  + \left( 1 - \frac{c}{b} \right) \log \frac{1-\frac{c}{b}}{1-\frac{b}{a}} \right]
    \notag \\ &  \quad \quad  +
(a-b) \left[ \left( 1 - \frac{b-c}{a-b} \right)  \log \frac{1 - \frac{b-c}{a-b}}{1- \frac{b}{a}} + \frac{b-c}{a-b} \log \frac{\frac{b-c}{a-b}}{\frac{b}{a}}  \right] 
    \\ & \quad = 
a \left[ \frac{b}{a} \mathbb{D}_2\left( \frac{c}{b} \middle| \middle| \frac{b}{a} \right) + \left( 1 - \frac{b}{a} \right) \mathbb{D}_2 \left( \frac{b-c}{a-b} \middle| \middle| \frac{b}{a} \right) \right] 
\end{align}
and thus proving the lemma.

\end{IEEEproof}

\subsection{Proof of Theorem~\ref{thm:codevid19}}
\begin{IEEEproof}

Once again $m = \times_{k \in \mcf{K}} m_k$ and $\mcf{M} = \times_{k \in \mcf{K}} \mcf{M}_k$.

The theorem will be proven by showing that Code Construction~\ref{cc:overlay} can produce $(r,\mcf{K},\gamma)$-overlay codes with non-zero probability.
Note, the fact that it can produce a code with non-zero probability directly implies the existence of such a code. 
Also note that the code construction near directly provides two of the $(r,\mcf{K}, \gamma)$-code requirements.
Indeed,
\begin{align}
\log |\mcf{M}| 
    &= 
\sum_{ k \in \mcf{K}}\log |\mcf{M}_k| 
    \\ &\geq 
\sum_{k \in \mcf{K} } n_k \left| \mathbb{I}_2\left(\gamma \middle|\middle| \frac{\ell}{n_k} \right) - \frac{4}{3n_k} - \frac{2}{n_k}\log  n_k \sqrt{\ell}  \right|^+
    \\ &\geq 
r
\end{align}
since $\left \lfloor e^{|a|^+}  \right \rfloor \geq e^{|a-1|^+}$.
While 
\begin{equation} 
\sum_{j=1}^n \idc{ f_j(m)}{\{ k \} } = \ell
\end{equation}
for each $k \in \mcf{K}$ and $m  \in \mcf{M}$ is already directly implied by the code construction.
What therefore remains to prove is that for each distinct pair of messages $m,m'$ there exists a $j\in \mcf{K}$ such that 
\begin{equation}\label{eq:codevid19:2}
\sum_{i=1}^n \idc{ f_i(m)}{\{ j  \} } \idc{f_i(m')}{\{j\}} \leq \gamma \ell 
\end{equation}
and for all $t < j$
\begin{equation}\label{eq:codevid19:2b}
\sum_{i=1}^n \idc{ f_i(m)}{\{ t  \} } \idc{f_i(m')}{\{t\}} =  \ell ,
\end{equation}
where it should be noted that Equation~\eqref{eq:codevid19:2b} implies
\begin{equation}
\sum_{i=1}^n \idc{ f_i(m)}{\{ k  \} } \idc{f_i(m')}{\{t\}} =  \ell .
\end{equation}
for all $t < j.$

To this end consider any $m = \times_{k \in \mcf{K}} m_k\in \mcf{M}$ and $m'  = \times_{k \in \mcf{K}} m_k' \in \mcf{M}$ such that $m \neq m'.$
Specifically, let\footnote{The value $j$ is used here since it will become the value of $j$ that satisfies Equations~\eqref{eq:codevid19:2} and~\eqref{eq:codevid19:2b}.} $j \in \mcf{K}$ be the minimum value such that $m_{k} \neq m_{j}'$, and note that for all $ t < j$ if $f_i(m) =  t$ then $f_i(m') = t$ since 
\begin{align}
& g_{\mcf{N}_t \rightarrow \mcf{I}\left(\times_{i \in \mcf{K}| i < t } m_i\right)}(\mcf{S}_{ t}(m_{t})) 
\notag \\ &\quad 
    = 
g_{\mcf{N}_t \rightarrow \mcf{I}\left(\times_{i \in \mcf{K}| i < t } m_i'\right)}(\mcf{S}_{k}(m'_{t}))
    \notag.
\end{align}
Clearly then for all $t < j$
\begin{equation}
\sum_{i=1}^n \idc{ f_i(m)}{\{ t  \} } \idc{f_i(m')}{\{t\}} =  \ell .
\end{equation}
What remains is to show that 
\begin{equation} 
\sum_{i=1}^n \idc{ f_i(m)}{\{ j  \} } \idc{f_i(m')}{\{j\}}  \leq \gamma \ell ,
\end{equation} 
which will be done via random coding arguments.
In particular, for each $k \in \mcf{K}$ we will show that with probability greater than zero the random choice of subsets in Code Construction~\ref{cc:overlay} yields a code such that 
\begin{equation}\label{eq:overlay:0}
|\mcf{S}_{k}(a) \cap \mcf{S}_k(b)| \leq \gamma \ell
\end{equation}
for all $k \in \mcf{K}$, $a \in \mcf{M}_k$, and $b \in \mcf{M}_k \setminus \{a\}.$
If a code with property~\eqref{eq:overlay:0} is produced, then 
\begin{equation} \label{eq:overlay:1}
\sum_{i=1}^n \idc{ f_i(m)}{\{ j  \} } \idc{f_i(m')}{\{j\}}  \leq |\mcf{S}_{j}(m_j) \cap \mcf{S}_j(m_j')| \leq \gamma \ell, 
\end{equation}
since the combination of $g_{\mcf{N}_j \rightarrow \mcf{I}\left( \times_{i\in \mcf{K}|i < j} m_i\right)} $ being an invertible mapping and 
\begin{align}
g_{\mcf{N}_j \rightarrow \mcf{I}\left( \times_{i\in \mcf{K}|i < j} m_i\right) } &= g_{\mcf{N}_j\rightarrow  \mcf{I}\left( \times_{i\in \mcf{K}|i < j} m_i'\right)}
    \notag.
\end{align}
imply that 
\begin{align}
&f_i(m) = f_i(m') = j 
    \notag \\ &\quad \Leftrightarrow 
g^{-1}_{\mcf{N}_j \rightarrow \mcf{I}\left( \times_{a\in \mcf{K}|a < j} m_i\right) } (\{i\}) \in  \mcf{S}_j(m_j) \cap \mcf{S}_j(m_j').
    \notag
\end{align}

To prove a code with Property~\eqref{eq:overlay:0} can be produced from the code construction, consider any $k \in \mcf{K}$, and without loss of generality assume $\mcf{M}_k = \{1,2,\dots,|\mcf{M}_k|\}.$
\com{\eric{WLOG because $\mcf{M}_k$ need not be defined, it could be a set containing all emojis for instance.}}
Further let $S_k(a)$ be the random variable representing the randomly chosen subset of $\mcf{N}_k$ particular to each $a\in\mcf{M}_k$.
Observe that the probability Code Construction~\ref{cc:overlay} generates a code satisfying~\eqref{eq:overlay:0} is 
\com{\eric{Using the notation of $\Pr (\text{event})$ here, I think it is more clear than setting everything up via indicators. Regardless it is messy.}}
\begin{align}
&\Pr \left( \cap_{a=2}^{|\mcf{M}_k|}   Q_{k}(a)   \right)
    =
\prod_{a=2}^{|\mcf{M}_k|} \Pr \left(  Q_k(a) \middle|  \cap_{b=1}^{a-1}  Q_k(b)  \right) 
    \label{eq:overlay:q}
\end{align}
where for each $a \in \mcf{M}_k$
$$Q_k(a) =  \bigcap_{b=1}^{a-1}
\{ |S_k(a)\cap S_k(b)| \leq \gamma \ell \}.$$
But, 
\begin{align}
&\Pr \left(  Q_k(a) \middle|  \cap_{b=1}^{a-1}  Q_k(b)  \right) 
    \notag \\ &\quad = 
1 -  \Pr \left( \cup_{c=1}^{a-1} |S_k(a)\cap S_k(c)| > \gamma \ell  \middle| \cap_{b=1}^{a-1}  Q_k(b)   \right) , 
    \label{eq:overlay:babykateisgangsta}\\ &\quad \geq 
1 -  \sum_{c=1}^{a-1} \Pr \left( |S_k(a)\cap S_k(c)| > \gamma \ell  \middle| \cap_{b=1}^{a-1}  Q_k(b)   \right) 
    \label{eq:overlay:babykateisgangsta2}\\ &\quad \geq 
1 -  (a-1) \Pr \left( |S_k(a)\cap S_k(1)| > \gamma \ell   \right) 
    \label{eq:overlay:babykateisgangsta3} ;
\end{align}
where~\eqref{eq:overlay:babykateisgangsta} follows by De Morgan's Law;~\eqref{eq:overlay:babykateisgangsta2} is the union bound; and~\eqref{eq:overlay:babykateisgangsta3} is because $\{S_k(a)\}_{a \in \mcf{M}_k}$ are independent and identically distributed.
Therefore, from combining equations~\eqref{eq:overlay:q},~\eqref{eq:overlay:babykateisgangsta3}, and the independence of the layer construction, it follows that if 
\begin{equation} \label{eq:overlay:nts}
\log|\mcf{M}_k| + \log \Pr \left( |S_k(1)\cap \mcf{S}_k(2)| > \gamma \ell    \right) < 0
\end{equation}
for all $k \in \mcf{K}$, then the probability Code Construction~\ref{cc:overlay} produces a code with property~\eqref{eq:overlay:0} for all values of $k\in \mcf{K}$ is greater than $0$. 
To prove Equation~\eqref{eq:overlay:nts} is indeed true, observe that 
\begin{align}
    & 
\log \Pr \left( |S_k(1)\cap \mcf{S}_k(2)| > \gamma \ell  \right) 
    \notag \\ & \quad 
    \geq 
\log \sum_{i=\lceil \gamma \ell \rceil}^{\ell} \Pr \left( |S_k(1)\cap \mcf{S}_k(2)| = i   \right) 
    \\ & \quad 
    = 
\log \sum_{i=\lceil \gamma \ell \rceil}^{\ell}  \frac{ \left| \left(\begin{matrix} \mcf{S}_k(2) \\ i \end{matrix} \right)\right| \left| \left(\begin{matrix} \mcf{N}_k - \mcf{S}_k(2) \\ \ell -i  \end{matrix} \right) \right| }{\left|\left(\begin{matrix}\mcf{N}_k \\ \ell \end{matrix} \right) \right|}
    \\ & \quad 
    = 
\log \sum_{i=\lceil \gamma \ell \rceil}^{\ell} \frac{\left(\begin{matrix} \ell \\ i \end{matrix} \right)\left(\begin{matrix} n_k - \ell \\ \ell -i  \end{matrix} \right) }{\left(\begin{matrix} n_k \\ \ell \end{matrix} \right)}
     \\ &\quad 
     \leq 
\log \max \left(  \sum_{i=\lceil \gamma \ell  \rceil}^{\ell} e^{-n_k \left( \mathbb{I}_2\left( \frac{i}{\ell} \middle|\middle| \frac{\ell}{n_k} \right) - \frac{1}{3n_k} - \frac{2}{n_k}\log n_k \right)}, 1 \right)
     \label{eq:overlay:yay}\\ &\quad 
     \leq 
-n_k  \left| \mathbb{I}_2\left(\gamma  \middle|\middle| \frac{\ell}{n_k} \right) - \frac{1}{3n_k} - \frac{2}{n_k}\log n_k \sqrt{\ell} \right|^+ 
    \label{eq:overlay:yay2} \\ &\quad 
    <
-\log|\mcf{M}_k|
    \label{eq:overlay:yay3};
\end{align}
where~\eqref{eq:overlay:yay} is from Lemma~\ref{lem:tec_codevid19} and because the probability of an event is at most $1$;~\eqref{eq:overlay:yay2} is because $\gamma \geq \frac{1}{2} \geq \frac{\ell}{n_k}$ for all $k$, in turn implying $$\mathbb{I}_2\left( \frac{i}{\ell} \middle|\middle| \frac{\ell}{n_k} \right) \geq \mathbb{I}_2\left( \frac{\lceil \gamma\ell \rceil}{\ell} \middle|\middle| \frac{\ell}{n_k} \right) \geq \mathbb{I}_2\left(  \gamma \middle|\middle| \frac{\ell}{n_k} \right) $$
and because a summation is always less than the maximum summand multiplied by total number of summands;
finally~\eqref{eq:overlay:yay3} is by code construction.
\end{IEEEproof}

\subsection{Proof of Corollary~\ref{cor:wedontwantthemtogetofftheshipbecausethatmaydoubleournumbers}}

\begin{IEEEproof}

Given $\gamma \in (\frac{1}{2},1)$ and finite $\mcf{K} \subset [0,1)$, recall that for all $k\in \mcf{K}$ 
$$n_k = n - j_k \ell  $$
where
$$  \ell=  \left \lfloor \frac{n}{|\mcf{\tilde K}|} \right\rfloor \quad \text{ and } \quad  j_k = |\{ a \in \mcf{K}| a < k\}|.$$

As a first step, observe that $\ell n_k \leq n^2$, hence
\begin{align}
&\sum_{k \in \mcf{K}} n_k \left| \mathbb{I}_2\left(\gamma  \middle|\middle| \frac{\ell}{n_k} \right) - \frac{4}{3n_k} - \frac{2}{n_k}\log \ell n_k  \right|^+
    \notag \\ &\quad \geq 
-|\mcf{K}| \left( \frac{4}{3} + 4 \log n \right) + \sum_{k \in \mcf{K}} n_k \mathbb{ I}_2\left(\gamma  \middle|\middle| \frac{\ell}{n_k} \right) \label{eq:wedontwantthemtogetofftheshipbecausethatmaydoubleournumbers:passiveagressiverefnames}
\end{align}
and what remains is to lower bound $n_k \mathbb{I}_2\left( \gamma \middle|\middle|\frac{\ell}{n_k} \right).$
To this end observe that
\begin{align}
&n_k \mathbb{I}_2\left( \gamma \middle|\middle|\frac{\ell}{n_k} \right)
    \notag \\ &\quad \geq 
\ell  \mathbb{D}_2\left( \gamma \middle|\middle|\frac{\ell}{n_k} \right)
        \label{eq:wedontwantthemtogetofftheshipbecausethatmaydoubleournumbers:42:1} \\ 
    & \quad \geq 
\ell \left[ \gamma \log \left(  |\mcf{\tilde K}|- j_k \right) + (1-\gamma) \log \left( \frac{n - j_k  \ell }{n-(j_k+1) \ell} \right) \right]  
    \notag \\ & \quad \quad - 
\ell \mathbb{H}_2(\gamma)
        \label{eq:wedontwantthemtogetofftheshipbecausethatmaydoubleournumbers:42:2}
 \\ & \quad \geq 
\ell \gamma \log \left(  |\mcf{\tilde K}|- j_k \right)  - \ell \mathbb{H}_2(\gamma)
    \label{eq:wedontwantthemtogetofftheshipbecausethatmaydoubleournumbers:42};
\end{align}
where~\eqref{eq:wedontwantthemtogetofftheshipbecausethatmaydoubleournumbers:42:1} is by the fact that the KL divergence is always greater than zero;~\eqref{eq:wedontwantthemtogetofftheshipbecausethatmaydoubleournumbers:42:2} is because 
$$\log \left( \frac{ n}{\left\lceil \frac{n}{|\mcf{\tilde K}|} \right\rceil} - j_k \right) \geq \log( |\mcf{\tilde K}| - j_k) ;$$ 
and~\eqref{eq:wedontwantthemtogetofftheshipbecausethatmaydoubleournumbers:42} is because $$n-j_k \ell \geq n-(j_k+1)\ell \geq 0.$$
Using this bound, 
\begin{align}
&  \sum_{k \in \mcf{K}} n_k \mathbb{ I}_2\left(\gamma  \middle|\middle| \frac{\ell}{n_k} \right) 
     \\ &\quad \geq
 \left \lfloor \frac{n}{|\mcf{\tilde K}|} \right \rfloor  \left( - |\mcf{K}| \mathbb{H}_2(\gamma) + \gamma \sum_{j = 0}^{|\mcf{K}|-1}  \log (|\mcf{\tilde K}|-j) \right)
     \\ &\quad =
\left \lfloor \frac{n}{|\mcf{\tilde K}|} \right \rfloor \left( - |\mcf{K}| \mathbb{H}_2(\gamma) + \gamma  \log (|\mcf{\tilde K}|)! \right)
     \\ &\quad \geq
 \left \lfloor \frac{n}{|\mcf{\tilde K}|} \right \rfloor \left( - |\mcf{K}| \mathbb{H}_2(\gamma) + \gamma  |\mcf{\tilde K}| \log \frac{|\mcf{\tilde K}|}{e} \right)
    \label{eq:wedontwantthemtogetofftheshipbecausethatmaydoubleournumbers:100k} \\ & \quad \geq
\left( n - |\mcf{K}| -1 \right) \left[ \gamma \log |\mcf{\tilde K}| - \gamma - \mathbb{H}_2(\gamma) \right],
\end{align}
where~\eqref{eq:wedontwantthemtogetofftheshipbecausethatmaydoubleournumbers:100k} is a consequence of Stirling's Approximation of the factorial.
Combining Equations~\eqref{eq:wedontwantthemtogetofftheshipbecausethatmaydoubleournumbers:passiveagressiverefnames},~\eqref{eq:wedontwantthemtogetofftheshipbecausethatmaydoubleournumbers:100k} with some further basic algebra and Theorem~\ref{thm:codevid19} yields the corollary statement.

\end{IEEEproof}

\section{Theorem~\ref{thm:1stcode}}\label{app:1stcode}

A technical lemma (which is essentially the Hoeffding lemma~\cite{hoeffding1963probability}), a basic calculation, an intuitively obvious lemma, and a bookkeeping lemma will be useful in the proof of Theorem~\ref{thm:1stcode}.
These are presented first, as to help streamline the proofs of the theorem.

\begin{lemma}\label{lem:hoeffck}
$$\Pr \left( \sum_{i=1}^{n}  G_{\rho,i}^2 \gtrless n (1\pm c)\rho \right) \leq \begin{cases} e^{-\frac{1}{8}c^2 n  }  &\text{if } c \leq 1 \\ e^{-\frac{1}{8} c n} &\text{else} \end{cases} $$
for all $c \geq 0.$
\end{lemma}
\begin{IEEEproof}

What follows is essentially the derivation of Hoeffding~\cite[Equation~(2.1)]{hoeffding1963probability} 
followed by a loosening of the bound. 
Proving $$\Pr \left( \sum_{i=1}^{n}  G_{\rho,i}^2 > n (1+c)\rho \right) \leq  e^{-\frac{1}{8}nc^2}$$
for all $c \geq 0 $ since a bound for  $\Pr \left( -\sum_{i=1}^{n}  G_{\rho,i} > n(1-c)\rho \right) $ follows with the same steps. 
Thus, the lemma can be derived as follows
\begin{align}
&\Pr \left( \sum_{i=1}^{n}  G_{\rho,i}^2 > n(1+c)\rho \right) 
    \notag \\&\quad = 
\min_{t >0 }  \Pr \left( e^{t \sum_{i=1}^{n}  G_{\rho,i}^2} > e^{tn(1+c)\rho}\right) \label{eq:hoeffck:bern} 
    \\&\quad \leq 
\min_{t >0 }  e^{-t n (1+c)\rho}  \mathbb{E}  e^{t \sum_{i=1}^{n}  G_{\rho,i}^2}  \label{eq:hoeffck:MI} 
    \\&\quad =  
\min_{t >0 }  e^{-tn(1+c)\rho} \prod_{i=1}^n \frac{1}{\sqrt{1-2t\rho}} \label{eq:hoeffck:MI2} 
    \\&\quad =  
\min_{t >0 }  e^{-tn(1+c)\rho }  \left(1-2t\rho\right)^{\frac{-n}{2}}
    \\&\quad =  
e^{-\frac{nc}{2}} \left(1+c \right)^{\frac{n}{2}}
 \label{eq:hoeffck:MI3}
    \\&\quad \leq  
\begin{cases} e^{-\frac{1}{8}c^2 n  }  &\text{if } c \leq 1 \\ e^{-\frac{1}{8} c n} &\text{else} \end{cases}\label{eq:hoeffck:fin}  
\end{align}
where~\eqref{eq:hoeffck:bern} is Bernstein's trick;
\eqref{eq:hoeffck:MI} is Markov's inequality;
\eqref{eq:hoeffck:MI2} is because $G_{\rho,i}$ is independent for each $i$ and hence
$$\mathbb{E}  e^{t \sum_{i=1}^{n}  G_{\rho,i}^2}  = \prod_{i=1}^{n} \mathbb{E} e^{t G_{\rho,i}^2} $$
while
\begin{align}
\mathbb{E}\left[e^{t G_{\rho,i}^2}\right]  &= \int_{\mcf{R}} \frac{1}{\sqrt{2\pi\rho}} e^{-\frac{x^2}{2\rho} + t x^2} \dx x 
    \notag\\&= 
\int_{\mcf{R}} \frac{1}{\sqrt{2\pi\rho}} e^{-\left(1- t2\rho \right) \frac{x^2}{2\rho}} \dx x 
    \notag\\&=
\frac{1}{\sqrt{1-2t\rho}} \int_{\mcf{R}} \frac{1}{\sqrt{2\pi\tilde \rho}} e^{- \frac{x^2}{2\tilde \rho}} \dx x     
    \notag\\&=
\frac{1}{\sqrt{1-2t\rho}} \notag
\end{align} 
where $\tilde \rho = \frac{\rho}{1-2t\rho};$
\eqref{eq:hoeffck:MI3} is the result of solving the minimization problem, and then substituting the minimum $t = \frac{c}{2(1+c)\rho}$ back in;
finally~\eqref{eq:hoeffck:fin} is because
$$ c - \log (1+c) \geq   \min_{c_0 \in [0,c]} \frac{c^2}{2} \frac{1}{1+c_0} \geq \begin{cases} \frac{1}{4} c^2  &\text{if } c \leq 1 \\ \frac{1}{4} c &\text{else} \end{cases} $$
for all $c \geq 0 $ by Taylor's theorem.

\end{IEEEproof}

\begin{calc}\label{calc:gauss}
If $X = G_{\rho}$ then
$$X|\{ X+ G_{a} = z\} = G_{\frac{\rho a}{\rho + a}} + \frac{\rho}{\rho+a} z
. $$
\end{calc}
\begin{IEEEproof}
Letting $Y = G_{a}$ and $Z= X + Y$, and $f_{X,Y,Z}$ denote the probability density functions of the various random variables, the calculation follows
\begin{align}
&f_{X|Z}(x|z) 
    \notag \\ & \quad = 
\frac{f_{Z|X}(z|x)f_{X}(x)}{f_{Z}(z)}
    \notag \\ &\quad= 
\sqrt{\frac{\rho + a}{2\pi\rho a}} \exp \left( - \frac{(z-x)^2}{2a} - \frac{x^2}{2\rho} + \frac{z^2}{2(\rho+a)} \right)
    \notag \\  &\quad= 
\sqrt{\frac{\rho + a}{2\pi\rho a}} \exp \left( - \frac{ x^2 - 2 \frac{\rho}{a+ \rho} x z + \frac{\rho^2 }{(a+ \rho)^2} z^2  }{2\frac{\rho a}{a+ \rho}} \right)
   \notag \\  &\quad= 
\sqrt{\frac{1}{2\pi\frac{\rho a}{\rho + a}}} \exp \left( - \frac{ \left(x - \frac{\rho}{\rho+a} z \right)^2 }{2\frac{\rho a}{a+ \rho}} \right)
.
\end{align}

\end{IEEEproof}

\begin{lemma} \label{lem:replace25}
Let $\mbf{G}$ be independent (but not identical) Gaussian RVs with mean $0$ and finite (but otherwise arbitrary) variance, and let $\mbf{\mu} \in \mbcf{R}$ be fixed. 
For all fixed $a >0 $
$$\Pr \left( \sum_{i=1}^n \left(G_i + \mu_i \right)^2 \leq a \right) \leq \Pr \left( \sum_{i=1}^n  G_i^2 \leq a \right) . $$
\end{lemma}
\begin{IEEEproof}
To prove the lemma, we need to show 
\begin{equation}\label{eq:r25:1}
\Pr \left( \left( G_i + \mu_i \right)^2 \leq a \right) \leq \Pr \left(  G_i^2 \leq a \right)
\end{equation}
since the more general lemma will then follow from repeated use of the following observation that uses Equation~\eqref{eq:r25:1}
\begin{align} 
&\Pr \left( \sum_{i=1}^n \left(G_i + \mu_i \right)^2 \leq a \right) 
    \notag \\ & \quad 
    = 
\int \!\!\Pr \left((G_1+\mu_1)^2 \leq a-b \right) \dx \Pr \left( \sum_{i=2}^n \left(G_i + \mu_i \right)^2 \leq b  \right)  
    \notag \\ & \quad 
    \leq 
\int \!\!\Pr \left(G_1^2 \leq a-b \right) \dx \Pr \left( \sum_{i=2}^n \left(G_i + \mu_i \right)^2 \leq b  \right)  
    \notag \\ & \quad 
    = 
\Pr \left( G_1^2+ \sum_{i=2}^n \left(G_i + \mu_i \right)^2 \leq a  \right)  .
    \notag 
\end{align}

To prove Equation~\eqref{eq:r25:1}, it is helpful to simplify it to
\begin{align}
&\Pr \left( - \mu_i - \sqrt{a} \leq G_i \leq -\mu_i + \sqrt{a} \right) 
    \notag \\ & \quad 
    \leq 
\Pr \left(  -\sqrt{a} \leq G_i \leq \sqrt{a} \right), \label{eq:r25:2}
\end{align}
or even more directly
\begin{align}
&\Phi\left( \frac{-\mu_i + \sqrt{a}}{\sqrt{\rho_i}} \right) - \Phi\left( \frac{-\mu_i - \sqrt{a}}{\sqrt{\rho_i}} \right) 
    \notag \\ & \quad 
\leq \Phi\left( \sqrt{\frac{a}{\rho_i}} \right) - \Phi\left( -\sqrt{\frac{a}{\rho_i}} \right),\label{eq:r25:3}
\end{align}
by taking square roots and then using basic algebraic manipulation.
Equation~\eqref{eq:r25:3} can be validated by showing that $\mu_i = 0 $ maximizes 
\begin{equation} \label{eq:r25:4}
\Phi\left( \frac{-\mu_i + \sqrt{a}}{\sqrt{\rho_i}} \right) - \Phi\left( \frac{-\mu_i - \sqrt{a}}{\sqrt{\rho_i}} \right)  .
\end{equation}
Using the basic calculus approach, the derivative of Equation~\eqref{eq:r25:4} is
\begin{equation} \label{eq:r25:5}
\frac{\partial \eqref{eq:r25:4}}{\partial \mu_i} = \frac{-1}{\sqrt{2\pi \rho_i} } \left[ e^{- \frac{(-\mu_i + \sqrt{a})^2}{2\rho_i}} - e^{- \frac{(-\mu_i - \sqrt{a})^2}{2\rho_i}} \right]  .
\end{equation}
Setting the derivative equal to zero and solving gives $|\mu_i + \sqrt{a}| =|-\mu_i + \sqrt{a}|,$ which can be further simplified to $2\mu = 0$ since $a>0.$
Furthermore, the second derivative at $\mu_i = 0$ is 
$$-\sqrt{\frac{2 a}{\pi \rho_i^3}} e^{-\frac{a}{2\rho_i}} <0 ,$$
thus guaranteeing that $\mu_i = 0$ is the global maximum in turn proving Equation~\eqref{eq:r25:1} and the lemma.

\end{IEEEproof}

\begin{lemma}\label{lem:iwish}
Suppose that $ \tau_i \geq \alpha > 0  $ for $i\in \{1,\dots, n\}$, and that $\beta \geq \alpha.$
Then for all positive real numbers $b,~c,$ and $\gamma$, where $ \gamma \leq \frac{1}{n} |\{ i \in \{1,\dots,n\} | \tau_i < \beta\}|$,
$$\Pr \left( \sum_{i=1}^n G_{\mbf{\tau},i}^2 \leq n(1+c)b \right) \leq e^{- \frac{1}{8} n \gamma  \lambda^2} + e^{- \frac{1}{8} n (1-\gamma)  \lambda^2}, $$
where 
$$\lambda = \max \left( 0 ,  1-\frac{(1+c) b}{\gamma \alpha + (1-\gamma) \beta} \right).$$
\end{lemma}
\begin{IEEEproof}
Choose any $\mcf{B} \subset \{1,\dots, n\}$ such that $|\mcf{B}| = n \gamma  $ and all coordinates in $\mcf{B}$ correspond to $\tau_i  < \beta$, i.e., 
$$\mcf{B} \subseteq \{ i \in \{1,\dots,n\} | \tau_i < \beta \}.$$
Let $\mcf{\bar B} = \{1,\dots,n\} \setminus \mcf{B}$.

Now the proof is trivial for $\lambda = 0$, otherwise when $\lambda > 0$ the results follows as so.
\begin{align}
&\Pr \left( \sum_{i=1}^n G_{\mbf{\tau},i}^2 \leq n(1+c)b\right)
    \notag \\ & \quad 
    =
\Pr \left( \sum_{i \in \mcf{B}} G_{\mbf{\tau},i}^2 + \sum_{i \in \mcf{\bar B} } G_{\mbf{\tau},i}^2 \leq n(1+c)b\right)
    \label{eq:iwish:1} \\ & \quad
    \leq 
\Pr \left( \sum_{i \in \mcf{B}} \frac{\alpha}{\tau_i}G_{\mbf{\tau},i}^2 + \sum_{i \in \mcf{\bar B} } \frac{\beta}{\tau_i}G_{\mbf{\tau},i}^2 \leq n(1+c)b\right)
    \label{eq:iwish:2} \\ & \quad 
    \leq 
\Pr \left( \sum_{i \in \mcf{B}} \frac{\alpha}{\tau_i}G_{\mbf{\tau},i}^2  \leq n\gamma (1-\lambda)  \alpha \right) 
    \notag \\ & \quad \quad 
    + 
\Pr \left( \sum_{i\in \mcf{\bar B}} \frac{\beta}{\tau_i}G_{\mbf{\tau},i}^2 \leq n(1-\gamma)(1-\lambda) \beta  \right)
    \label{eq:iwish:3} \\ & \quad 
    \leq 
e^{-\frac{1}{8} n \gamma \lambda^2} + e^{-\frac{1}{8} n (1-\gamma) \lambda^2}
\end{align}
where~\eqref{eq:iwish:2} is because $\alpha/\tau_i \leq 1 $ for all $i \in \mcf{B}$ and $\beta/\tau_i \leq 1$ for all $i\in \mcf{\bar B}$;~\eqref{eq:iwish:3} is by using the inequality
\begin{align}
\Pr (A+B\leq a+b) &\leq \Pr (\{A\leq a\} \text{ or } \{B \leq b\}) \
    \notag \\ & 
    \leq 
    \Pr (A\leq a) + \Pr (B \leq b) 
    \notag 
\end{align}
in conjunction with
$$n(1+c)b \leq n \gamma (1 - \lambda) \alpha + n(1-\gamma) (1-\lambda) \beta; $$
and~\eqref{eq:iwish:3} is by Lemma~\ref{lem:hoeffck} and because $\frac{\alpha}{\tau_i} G_{\mbf{\tau},i}^2 = G_{\alpha}^2$ and  $\frac{\beta}{\tau_i} G_{\mbf{\tau},i}^2 = G_{\beta}^2.$
\end{IEEEproof}
\subsection{Proof of Theorem~\ref{thm:1stcode}}

\begin{IEEEproof}
Let $\mcf{H} = (\mbf{x}: \mcf{M} \rightarrow \mbcf{R}$, $\hat m' : \mbcf{R} \rightarrow \mcf{M})$ be the original code, and let $\mcf{J} = ( \mbf{X}' : \mcf{M} \rightarrow \mbcf{R}$, $\hat m' : \mbcf{R} \rightarrow \mcf{M} \cup \{!\})$ be the modified code obtained from Code Modification~\ref{code:addnoise}.
By $\mcf{I}_k(m)$, for each $k \in \mcf{\tilde K}$ and $m \in \mcf{M},$ denote all coordinates $i \in \{1,\dots,n\}$ such that $f_i(m) = k$.

Both the power constraint and the average arithmetic probability of error arguments will rely on random coding (due to the random choice of $\mbf{t}: \mcf{M} \rightarrow \mbcf{R}).$
Because of this, let $\mbf{T}: \mcf{M} \rightarrow \mbcf{R}$ be the random variable representing the randomly chosen value of $\mbf{t}$ in the code construction.
The random coding construction will proceed by showing that the random choice of $\mbf{T}$ with probability greater than $$1 -\left(1 + \sqrt{\frac{2}{\pi}} \right)e^{-n}$$ yields a code with the stated power constraint, and with probability greater than $$1-e^{-n}$$ yields a code with the stated average arithmetic probability of error. 
Clearly, this also implies that the random choice of $\mbf{T}$ yields a code which satisfies both the power constraint and the average arithmetic probability of error bound with probability greater than $$1 -\left(2 + \sqrt{\frac{2}{\pi}} \right)e^{-n}.$$

For readability, we have separated the derivation of each bound by  a dividing line.  
~\\
\hrule
~\\
\emph{(Rate)}~\\
\indent Encoders for $\mcf{J}$ and $\mcf{H}$ have the same domain hence
$$r_{\mcf{J}} = r_{\mcf{H}}.$$
\hrule
~\\
\emph{(Power)}~\\
\indent Towards the power constraint observe that for each message $m\in \mcf{M}$
\begin{align}
&\sum_{i=1}^n \frac{1}{n} \mathbb{E}\left[ ( X_i'(m))^2 \right] 
    \notag \\ &\quad   = 
\sum_{i=1}^n \frac{1}{n} \mathbb{E}\left[ ( x_i(m)  +t_i(m)+ f_i(m)  G_{\Delta,i}) ^2 \right] 
    \\ &\quad     = 
\sum_{i=1}^n \frac{1}{n} \left(   x^2_i(m)  +t_i^2(m)  + 2t_i(m) x_i(m) +  f_i^2(m) \rho_{\Delta}  \right) ;\label{eq:1stcode:powerbound0:1}
    \\ &\quad     \leq 
\omega_{\mcf{H}} + \frac{1}{n} \sum_{i=1}^n 2 t_i x_i +\frac{1}{n} \sum_{k\in \mcf{\tilde K}} \sum_{i \in \mcf{I}_k(m)}  \left(  t_i^2(m)  +  k^2 \rho_{\Delta}  \right) ;\label{eq:1stcode:powerbound0}
\end{align}
where~\eqref{eq:1stcode:powerbound0:1} is because $\mathbb{E}\left[ G_{\Delta,i} ^2 \right] = \rho_{\Delta}$ and $\mathbb{E}\left[ G_{\Delta,i} \right]=0;$ and~\eqref{eq:1stcode:powerbound0} is by definition of the power constraint.
Thus, we need to bound the tail probability for choosing large values of $t_i^2(m)$ and $2t_i(m) x_i(m).$

To this end, for each $m\in \mcf{M}$ and $k \in \mcf{K}$ 
\begin{align}
    &
\Pr \left( \sum_{i \in \mcf{I}_k(m)} \hspace{-7pt} T_i^2(m) \geq \ell \left( 1 + c \right) (1-k^2)\rho_{\Delta} \right)
    \notag \\ & \quad 
    \leq 
e^{-n \left[ r_{\mcf{H}} +1\right] - \log |\mcf{ K}| }  \label{eq:1stcdoe:powerbound1},
\end{align}
where $$c = 8 \frac{n}{\ell}\left[ r_{\mcf{H}} + 1+ \frac{\log{|\mcf{K}|}}{n} \right]> 1, $$ by Lemma~\ref{lem:hoeffck}. 
That with probability $1 - e^{-n}$ a $\mbf{T} = \mbf{t}$ is chosen such that
\begin{align}
    &
 \sum_{i =1}^n (t_i^2 (m) + k^2 \rho_{\Delta} ) 
    \leq 
n \left( 1 + 8 \frac{n}{\ell}\left[ r_{\mcf{H}} + 1+ \frac{\log{|\mcf{K}|}}{n} \right] \right) \rho_{\Delta}  \label{eq:1stcode:powerbound:wtf1}
\end{align} 
for all $m$ follows by applying the union bound to extend~\eqref{eq:1stcdoe:powerbound1} to simultaneously consider all $m\in \mcf{M}$ and $k \in \mcf{K}$ (while observing that $t_i(m) = 0$ for all $i \in \mcf{I}_1(m)$).

The other term in the summation, $\sum_{i=1}^n 2 x_i(m) T_i(m)$, follows directly from basic laws of probability.
Specifically, for each $m\in \mcf{M}$ we have
\begin{align}
    &
\Pr \left( \sum_{i=1}^n 2 x_i(m) T_i(m) \geq   n 2 \sqrt{ 2\omega_{\mcf{H}}(r_{\mcf{H}}+1) \rho_{\Delta}}\right)
    \notag \\ & \quad 
    = 
\Phi \left( \frac{-n 2\sqrt{2\omega_{\mcf{H}}(r_{\mcf{H}}+1) \rho_{\Delta}}}{\sqrt{  \sum_{i=1}^n 4 x_i^2(m)(1-f_i^2(m)) \rho_{\Delta}}} \right)
    \label{eq:1stcdoe:powerbound2:1} \\ & \quad  
    \leq 
\sqrt{\frac{2}{\pi}} \exp\left(- \frac{ n^2 \omega_{\mcf{H}} (r_{\mcf{H}} +1)}{  \sum_{i=1}^n  x_i^2(m)(1-f_i^2(m))} \right)
    \label{eq:1stcdoe:powerbound2:2} \\ & \quad 
    \leq 
\sqrt{\frac{2}{\pi}} e^{- n(r_{\mcf{H}} +1)} \label{eq:1stcdoe:powerbound2};
\end{align}
where~\eqref{eq:1stcdoe:powerbound2:1} is because $2x_i(m) T_i(m)$ is a sum of independent Gaussian random variables by the code construction;~\eqref{eq:1stcdoe:powerbound2:2} is because $\Phi(t) \leq \sqrt{\frac{2}{\pi}} e^{-\frac{t^2}{2}}$ for $ t\leq 0$; and~\eqref{eq:1stcdoe:powerbound2} is because $0\leq f_i(m) \leq 1 $ for all coordinates $i \in \{1,\dots,n\}$ and $m \in \mcf{M}.$
Once again, that with probability greater than $1- \sqrt{\frac{2}{\pi}} e^{-n}$ a $\mbf{T} = \mbf{t}$ is chosen such that 
\begin{equation} \label{eq:1stcode:powerbound:wtf2}
 \sum_{i =1}^n 2 x_i(m) t_i(m) \leq n 2 \sqrt{2 \omega_{\mcf{H}}( r_{\mcf{H}} +1) \rho_{\Delta}}
\end{equation}
follows by applying the union bound to Equation~\eqref{eq:1stcdoe:powerbound2} as to consider all $m$ jointly.

Combining Equations~\eqref{eq:1stcode:powerbound0},~\eqref{eq:1stcode:powerbound:wtf1},~\eqref{eq:1stcode:powerbound:wtf2}, and that $\frac{n}{\ell}\leq |\mcf{\tilde K}|$ shows that with probability
$$ 1 -\left( 1 + \sqrt{\frac{2}{\pi}} \right) e^{-n}  $$
a $\mbf{T} = \mbf{t}$ is chosen such that
\begin{align}
 \omega_{\mcf{J}}  &\leq \omega_{\mcf{H}} + 2\sqrt{ 2 \omega_{\mcf{H}}  (r_{\mcf{H}} + 1)  \rho_{\Delta}} 
    \notag \\ & \quad 
+   \left( 1 \!+ \!8 |\mcf{\tilde K}|  \left[ r_{\mcf{H}} \!+\!1\! + \!\frac{\log |\mcf{K}|}{n} \right] \right)\rho_{\Delta}.
\end{align}
\hrule
~\\
\emph{(Average arithmetic probability of error)}~\\
\indent To prove the bound on the average arithmetic error probability, observe that the condition for error given $M=m$, 
\begin{equation} 
\hat m'(\mbf{x}(m) + \mbf{t}(m) + \mbf{f}(m) \cdot \mbf{G}_{\Delta} + \mbf{G}_{\mathrm{Dec}} ) \neq m, 
\end{equation} 
occurs if and only 
\begin{equation} 
\hat m(\mbf{x}(m) + \mbf{t}(m) + \mbf{f}(m) \cdot \mbf{G}_{\Delta} + \mbf{G}_{\mathrm{Dec}} ) \neq m, 
\end{equation}
or if there exists a $k \in \mcf{K}$ such that
\begin{equation} 
\sum_{i \in \mcf{I}_{k}(m)} \frac{(k G_{\Delta,i} + G_{\mathrm{Dec},i})^2}{k^2\rho_{\Delta} + \rho_{\mathrm{Dec}}}  \geq \ell( 1+ \delta) .
\end{equation}
Hence
\begin{equation}
\varepsilon_{\mcf{J}} (\rho_{\mathrm{Dec}})   \leq \frac{1}{|\mcf{M}|} \sum_{m\in \mcf{M}}  a_{m}(\mbf{t}) + \frac{1}{|\mcf{M}|}\sum_{\substack{k\in \mcf{K}\\ m \in \mcf{M}}} b_m(k) ,
\end{equation}
where
\begin{align}
a_m(\mbf{t}) &= \Pr \left( \hat m(\mbf{x}(m) + \mbf{t}(m) + \mbf{f}(m) \cdot \mbf{G}_{\Delta} + \mbf{G}_{\mathrm{Dec}} ) \neq m \right) 
    \notag \\
b_m(k) &= \Pr \left( \sum_{i \in \mcf{I}_{k}(m)} \frac{(k G_{\Delta,i} + G_{\mathrm{Dec},i})^2}{k^2\rho_{\Delta} + \rho_{\mathrm{Dec}}}  \geq \ell( 1+ \delta) \right),
    \notag
\end{align}
by the union bound. 
Note from the code construction that random variables $\mbf{a}_{m}(\mbf{T})$ and $\mbf{a}_{m'}(\mbf{T})$ are independent when $m\neq m'$, and that $0 \leq a_{m}(\mbf{T}) \leq 1$ for all $m\in \mcf{M}$, and that
\begin{align}
\mathbb{E}[a_m(\mbf{T})] = \Pr \left( \hat m(\mbf{x}(m) + \mbf{G}_{\rho_{\Delta}+ \rho_{\mathrm{Dec}}} ) \neq m \right) 
\end{align}
because $T_i(m) + f_i(m) G_{\Delta,i} + G_{\mathrm{Dec},i}$ has a Gaussian distribution with mean $0$ and variance $\rho_{\Delta}+ \rho_{\mathrm{Dec}}$ for each coordinate $i$. 
Therefore
\begin{align}
\Pr \left(\sum_{m \in \mcf{M}} \frac{a_{m}(\mbf{T})}{|\mcf{M}|} \geq \varepsilon_{\mcf{H}}(\rho_{\Delta}+ \rho_{\mathrm{Dec}}) + \sqrt{\frac{n}{2|\mcf{M}|}} \right) 
\leq  e^{-n}
\end{align}
follows from Hoeffding's inequality because 
$$\sum_{m\in \mcf{M}}\frac{\Pr \left( \hat m(\mbf{x}(m) + \mbf{G}_{\rho_{\Delta}+ \rho_{\mathrm{Dec}}} ) \neq m \right) }{|\mcf{M}|} = \varepsilon_{\mcf{H}} (\rho_{\Delta}+ \rho_{\mathrm{Dec}}) .$$
On the other hand 
\begin{align}
b_{m}(k) \leq e^{-\frac{1}{8} \ell \delta^2} 
\end{align}
comes directly from Lemma~\ref{lem:hoeffck} since $\frac{kG_{\Delta,i} + G_{\mathrm{Dec},i}}{\sqrt{k^2\rho_{\Delta}+\rho_{\mathrm{Dec}}}}$ are independent Gaussian random variables with mean $0$ and variance $1$ for each coordinate $i$.
Therefore, with probability greater than $1-e^{-n},$ a function $\mbf{T}=\mbf{t}$ will be chosen such that
\begin{align}
\varepsilon_{\mcf{J}}(\rho_{\mathrm{Dec}})  
\leq \varepsilon_{\mcf{H}}(\rho_{\mathrm{Dec}}+ \rho_{\Delta})  + \sqrt{\frac{n}{2|\mcf{M}|}} +  |\mcf{K}| e^{-\frac{1}{8} \ell \delta^2} \label{eq:1stcode:out1-1}.
\end{align}
\hrule
~\\
\emph{(Maximum probability of targeted false authentication)}~\\
\indent To prove the bound on the maximum probability of targeted false authentication, fix messages $M=m$ and $m'\neq m$ with the intention that $m'$ is the targeted message, let $k \in \mcf{K}$ be the index such that 
$$ \sum_{i=1}^n \idc{ f_i(m)}{\{ k  \} } \idc{f_i(m')}{\{k\}}  \leq \gamma \ell ,$$
and
$$\sum_{i=1}^n \idc{ f_i(m)}{\{ k  \} } \idc{f_i(m')}{\{j\}}  = 0$$
for all $j< k .$
The probability that the decoder will produce $m'$ is always less than
\begin{align}
\Pr \left( \sum_{i \in \mcf{I}_{k}(m')} \frac{(Y_i -t_i(m') - x_i(m'))^2}{k^2 \rho_{\Delta} + \rho_{\mathrm{Dec}}}   \leq \ell( 1+ \delta) \right) , \label{eq:1stcode:advgoal}
\end{align}
where as a reminder
\begin{align}
\mbf{Y} &= \mbf{x}(m)+ \mbf{t}(m) +\mbf{f}(m) \cdot \mbf{G}_{\Delta} + \mbf{Z}(\mbf{V},m) ,
    \notag 
\end{align}
due to the code modification to the decoder. 

Assume for now (we will come back to prove this after finishing the proof, see after break) that
\begin{equation}\label{eq:1stcode:advgoalEquiv}
\eqref{eq:1stcode:advgoal} \leq \Pr \left( \sum_{i\in \mcf{I}_k(m')}  G_{\mbf{\tau}(m),i}^2   \leq \ell (1+\delta)(k^2 \rho_{\Delta} + \rho_{\mathrm{Dec}}) \right) \hspace{-3pt},
\end{equation}
where 
\begin{align}
\tau_i(m) &=  \tau^\star (f_i(m)) := \frac{f_i^2(m)\rho_{\Delta} \rho_{\mathrm{Adv}} }{f_i^2(m)\rho_{\Delta} + \rho_{\mathrm{Adv}}} + \rho_{\mathrm{Dec}}.
    \notag
\end{align}
With Equation~\eqref{eq:1stcode:advgoal} assumed, the following properties of the overlay-code allow for application of Lemma~\ref{lem:iwish}: 
\begin{itemize}
    \item $$|\mcf{I}_{k}(m')| = \ell,$$
    \item $$|\mcf{I}_k(m') \cap \mcf{I}_k(m)| \leq \gamma \ell,$$
    \item $$ \tau_i(m) =  \tau^\star (k) = \frac{k^2 \rho_{\Delta} \rho_{\mathrm{Adv}}}{k^2 \rho_{\Delta} +  \rho_{\mathrm{Adv}}} + \rho_{\mathrm{Dec}} $$
for all $i \in \mcf{I}_k(m') \cap \mcf{I}_k(m)$,
    \item and $$\tau_i(m) \geq  \tau^\star(d_k) = \frac{d_k^2 \rho_{\Delta} \rho_{\mathrm{Adv}}}{d_k^2 \rho_{\Delta} +  \rho_{\mathrm{Adv}}} + \rho_{\mathrm{Dec}},$$ 
where $d_k = \min \{ a \in \mcf{\tilde K} | a > k\}$, for all $i \in \mcf{I}_k(m') \setminus \mcf{I}_k(m)$.
\end{itemize}
With these overlay-code properties we can directly apply Lemma~\ref{lem:iwish} to upper-bound the right-hand side of Equation~\eqref{eq:1stcode:advgoalEquiv} in turn yielding
\begin{equation}\label{eq:1stcode:advgoalfin}
\eqref{eq:1stcode:advgoal} \leq e^{-\frac{1}{8}\ell \gamma \lambda_k^2 } + e^{-\frac{1}{8}\ell (1- \gamma) \lambda_k^2 },
\end{equation}
where 
$$\lambda_k = \max \left( 0 , 1 - \frac{(1+\delta)(k^2 \rho_{\Delta} + \rho_{\mathrm{Dec}})}{\gamma \tau^\star(k) + (1-\gamma) \tau^\star(d_k)} \right).$$

Recall now that Equation~\eqref{eq:1stcode:advgoal} assumed a fixed $m$ and $m' \neq m$ and that the maximum probability of targeted false authentication is a maximum over all pairs of $m, m' \neq m.$
Thus the maximum, over all $m$ and $m' \neq m$, of the right-hand side of Equation~\eqref{eq:1stcode:advgoalfin} is also an upper bound on $\alpha^{\star}_{\mcf{H}}(\rho_{\mathrm{Adv}},\rho_{\mathrm{Dec}})$.
Clearly though, the maximum of the right-hand side of Equation~\eqref{eq:1stcode:advgoalfin} corresponds to the minimum value of $\lambda_k.$
Hence the final result
\begin{equation}\label{eq:1stcode:advgoalfin2}
\alpha^{\star}_{\mcf{H}}(\rho_{\mathrm{Adv}},\rho_{\mathrm{Dec}})  \leq e^{-\frac{1}{8}\ell \gamma \lambda^2 } + e^{-\frac{1}{8}\ell (1- \gamma) \lambda^2 },
\end{equation}
where 
$$\lambda = \max \left( 0 , \min_{k \in \mcf{K}} 1 - \frac{(1+\delta)(k^2 \rho_{\Delta} + \rho_{\mathrm{Dec}})}{\gamma \tau^\star(k) + (1-\gamma) \tau^\star(d_k)} \right).$$

~\\~\\~\\
\indent We now return to prove Equation~\eqref{eq:1stcode:advgoalEquiv}.
Here we will primarily use the inequality $\Pr \left( \cdot \right) \leq \sup_{a} \Pr \left( \cdot \middle| A = a \right) $ along with calculation~\ref{calc:gauss}. 
To that end note
\begin{align} 
&\mbf{Y} - \mbf{x}(m') - \mbf{t}(m')  |\{ \mbf{V},\mbf{Z}=\mbf{v},\mbf{z} \} 
    = 
\mbf{G}_{\mbf{\tau}(m)} + \mbf{\mu}(\mbf{v},\mbf{z})
\end{align}
where
\begin{align}
\mu_i(\mbf{v},\mbf{z}) &=   z_i - x_i(m') - t_i(m') 
    \notag \\ & \quad 
+ \frac{f^2_i(m)\rho_{\Delta} (v_i-x_i(m) - t_i(m) ) }{f_i^2(m)\rho_{\Delta} + \rho_{\mathrm{Adv}}} 
    \notag \\ 
\tau_i(m) &=  \frac{f_i^2(m)\rho_{\Delta} \rho_{\mathrm{Adv}} }{f_i^2(m)\rho_{\Delta} + \rho_{\mathrm{Adv}}} + \rho_{\mathrm{Dec}},
\notag
\end{align}
as a consequence of calculation~\ref{calc:gauss}.
Hence~\eqref{eq:1stcode:advgoal} must itself be less than 
\begin{align}\label{eq:1stcode:advgoal:goal}
\sup_{\mbf{v},\mbf{z}} \Pr \left( \sum_{i\in \mcf{I}_k(m')}  \left(G_{\mbf{\tau}(m),i} + \mu_i(\mbf{v},\mbf{z})\right)^2  \leq c \middle| \mbf{V},\mbf{Z} = \mbf{v},\mbf{z} \right) 
\end{align}
where $c = \ell(1+\delta) (k^2\rho_{\Delta} + \rho_{\mathrm{Dec}})$.
Applying Lemma~\ref{lem:replace25} to~\eqref{eq:1stcode:advgoal:goal}, and recognizing that the resulting probability is independent of $\mbf{V},\mbf{Z}$ proves
\begin{equation}
\eqref{eq:1stcode:advgoal} \leq \Pr \left( \sum_{i\in \mcf{I}_k(m')}  \left(G_{\mbf{\tau}(m),i}\right)^2  \leq c\right),
\end{equation}
which is exactly Equation~\eqref{eq:1stcode:advgoalEquiv}.

\end{IEEEproof}

\section{Theorem~\ref{thm:2ndcode}}\label{app:2ndcode}

The proof of the Theorem~\ref{app:2ndcode} will rely on the following technical lemmas.

The first of these technical lemmas will be used to create a finite subset of points which when bounded also bound the original set.

\begin{lemma}\label{lem:onade}
Let $\mcf{X}^* \subseteq [a,b]$, for real numbers $a$ and $b>a$. 

For a given positive real number $c$, there exists an $\mbcf{X}^\dagger \subset \mbcf{R}$ such that $|\mbcf{X}^\dagger| = \lfloor (b-a) c \rfloor^n $ and 
\begin{align} 
&\sup_{\mbf{x}^* \in \mbcf{X}^*} \Pr \left( \mbf{G}_{\mbf{\rho}} + \mbf{x}^* \in \mcf{D} \right) 
    \notag \\ & \quad
    \leq 
\max_{\mbf{x} \in \mbcf{X}^\dagger} \Pr \left( \mbf{G}_{\mbf{\rho}} + \mbf{x}^\dagger \in \mcf{D} \right)  + c^{-1}\sqrt{\sum_{i=1}^n \frac{1}{2\rho_i}}
    \notag 
\end{align}
simultaneously for all $\mcf{D} \subseteq \mbcf{R}$.

\end{lemma}

\begin{IEEEproof}

First we identify  $\mbcf{X}^\dagger$, where
\begin{align} 
&\mcf{X}^\dagger =  
\left\{a + c^{-1}, a + 2c^{-1}, \dots, a + \lfloor (b-a) c \rfloor c^{-1} \right\}\notag ,
\end{align}
as the set guaranteed in the lemma.
It is immediate that $|\mbcf{X}^\dagger| = \lfloor (b-a) c \rfloor^n$.

Now, for each $\mbf{x} \in \mbcf{X}^*$ consider the corresponding $\mbf{x}^\dagger= \argmin_{\mbf{x}^\dagger \in \mbcf{X}^\dagger} |\mbf{x} - \mbf{x}^\dagger|$, and note that $\mbf{x}^\dagger \in \mbcf{X}^\dagger$ by definition. 
Here, $|x_i - x_i^\dagger|\leq c^{-1}$ for all coordinates $i \in \{1,\dots,n\}$.
Hence, for $\mbf{x}$ and corresponding $\mbf{x}^\dagger$ it follows that
\begin{align}
    &
\left| \Pr \left( \mbf{G}_{\mbf{\rho}} + \mbf{x} \in \mcf{D} \right) - \Pr \left( \mbf{G}_{\mbf{\rho}} + \mbf{x}^\dagger \in \mcf{D} \right) \right| 
    \notag \\& \quad 
    \leq 
\sqrt{\frac{1}{2} \mathbb{D}(\mbf{G}_{\mbf{\rho}}+ \mbf{x}||\mbf{G}_{\mbf{\rho}} + \mbf{x}^\dagger)} 
     \label{eq:lemonade:2}\\ & \quad 
     \leq
c^{-1}\sqrt{\sum_{i=1}^n \frac{1}{2\rho_i}}; \label{eq:lemonade:3}
\end{align}
where~\eqref{eq:lemonade:2} is by Pinsker's inequality and the convexity of the KL divergence; while~\eqref{eq:lemonade:3} is because
\begin{align} 
\mathbb{D}\left( \mbf{G}_{\mbf{\rho}} + \mbf{x} \middle| \middle| \mbf{G}_{\mbf{\rho}} + \mbf{x}^\dagger \right) = \sum_{i=1}^n \frac{(x_i-x_i^\dagger)^2}{2 \rho_i} \leq c^{-2} \sum_{i=1}^n \frac{1}{2 \rho_i}. 
\notag 
\end{align}
This proves the lemma since for all $\mbf{x} \in \mbcf{X}^*$ there is a corresponding $\mbf{x}^\dagger \in \mbcf{X}^\dagger$ such that Equation~\eqref{eq:lemonade:3} holds independent of $\mcf{D}$.

\end{IEEEproof}

Next we provide a corollary of the well known Hoeffding Lemma.
While we will prove the corollary, we point readers to~\cite[Section~5]{hoeffding1963probability} for proof of the lemma, and note that uniformly selecting $m$ values without replacement is equivalent to uniformly selecting a subset of size $m$.
\begin{lemma}\label{lem:hoeffwo} \textbf{(\!\!\cite[Section~5]{hoeffding1963probability})} 
Let $\mcf{A}$ be a finite set, $\beta$ an integer less than $|\mcf{A}|$, and let $p: \mcf{A} \rightarrow [0,1]$.
If $B$ is uniform over $\left( \begin{matrix} \mcf{A} \\ \beta \end{matrix} \right)$
then
\begin{align} 
\Pr \left( \sum_{a \in \mcf{A}}\idc{a}{\{B\}}  p(a) \geq c  \beta  \mu \right) &\leq \exp (-\beta \mathbb{D}_2( c \mu || \mu)), 
    \notag \\
&\leq \exp (-2\beta [(c-1)\mu]^2), 
    \notag
\end{align}
where $\mu = |\mcf{A}|^{-1} \sum_{a \in \mcf{A}} p(a)$, for all real numbers $c \in (1 , \mu^{-1}) .$

\end{lemma}

\begin{cor}\label{cor:dcutting}
Additionally if $\max_{a \in \mcf{A}} p(a) = \eta < 1$ then 
\begin{align} 
&\Pr \left( \sum_{a \in \mcf{A}}\idc{a}{  \{B\}}  p(a) \geq c \beta \mu \right) 
    \notag \\ &\quad 
\leq \exp\left(-\beta \mathbb{D}_2\left(c \frac{ \mu}{\eta} \middle|\middle| \frac{\mu}{\eta} \right) \right) 
    \notag \\ &\quad \leq 
\exp \left( -\frac{c\beta \mu}{\eta} \left( \log (c)  -\frac{1}{2} - \frac{1}{2\left( 1 - c\frac{\mu}{\eta} \right)}  \right) \right) 
    \notag 
\end{align}
for all real numbers $c \in (1 ,\eta \mu^{-1}) .$
\end{cor}
\begin{IEEEproof}
The first inequality comes from substituting $\frac{p(a)}{\eta}$ for $p(a)$ (and subsequently $\frac{\mu}{\eta}$ for $\mu$) in Lemma~\ref{lem:hoeffwo}.

The second inequality comes from recognizing that $$\left(1-\frac{\mu}{\eta}c \right) \log \frac{1-\frac{\mu}{\eta}c}{1-\frac{\mu}{\eta}} \geq \left(1-\frac{\mu}{\eta}c \right) \log \left(1-\frac{\mu}{\eta}c\right) $$ and that if $a\in [0,b]$, where $0 \leq b\leq1 $, then $$(1-a) \log(1-a) \geq -a - \frac{a^2}{2(1-b)} \geq -a - \frac{a^2}{2(1-a)}$$ by Taylor's theorem.
\end{IEEEproof}

\subsection{Proof of Theorem~\ref{thm:2ndcode}}
\begin{IEEEproof}

Since Code Modification~\ref{code:decimate} builds on Code Modification~\ref{code:addnoise}, let $\mcf{H} = (\mbf{x}: \mcf{M} \rightarrow \mbcf{R}, \hat m : \mbcf{R} \rightarrow \mcf{M})$ be the original encoder and decoder, and let $\mcf{L} = ( \mbf{X}': \mcf{M} \rightarrow \mbcf{R} ,\hat m': \mbf{R} \rightarrow \mcf{M}\cup\{\mbf{!}\})$ be the code after Code Modification~\ref{code:addnoise}.
We will assume that the operational measures of $\mcf{L}$ are bounded as in Theorem~\ref{thm:1stcode}.

From the Proof of Theorem~\ref{thm:1stcode} it is important to recall that for each $m,\mbf{v},\mbf{z}$ there exists some $\mbf{\mu} \in \mbcf{R}$ such that
\begin{align} 
&\mbf{Y} |\{M,\mbf{V}, \mbf{Z} = m,\mbf{v},\mbf{z}\} 
    = 
\mbf{G}_{\mbf{\tau}(m)} + \mbf{\mu}
\end{align}
where 
$$\tau_i(m) = \tau^\star(f_i(m)) =  \frac{f_i^2(m)\rho_{\Delta} \rho_{\mathrm{Adv}} }{f_i^2(m)\rho_{\Delta} + \rho_{\mathrm{Adv}}} + \rho_{\mathrm{Dec}}.$$
Key to the proof of the upper-bound on the maximum probability of false authentication is that the decimation of the message set will not change the above. 

The proof will rely on the random selection of the new (decimated) message set, $\mcf{M}^\ddagger$, for the final code $\mcf{J} = (\mbf{X}^\ddagger : \mcf{M}^\ddagger \rightarrow \mbcf{R}, \hat m^\ddagger : \mbcf{R} \rightarrow \mcf{M}^\ddagger \cup \{ \mbf{!}\})$.
To represent this random selection, let $M^\ddagger$ be the random variable representing the chosen value of $\mcf{M}^\ddagger$ in the code construction of~\ref{code:decimate}. 
The random code construction will be useful in calculating bounds for both the average arithmetic error and the maximum probability of false authentication.
In particular, we will show that with probability 
$1-e^{-n}$
the randomly chosen value of $\mcf{M}^\ddagger$ yields a code with the stated average arithmetic error bound, and with probability 
$1-e^{-n/2}$
yields a code with the stated maximum probability of false authentication bound.
Note then the probability of selecting a code which satisfies both bound simultaneously must be at least $1-e^{-n}-e^{-n/2}$ due to the union bound.

Once again for readability, we have separated by a dividing line the bound for each for the operational measures. 
~\\
\hrule
~\\
\emph{(Rate)}~\\
\indent For the rate, first assume that $r_{\mcf{J}} \geq \frac{\log 2 n}{n} .$
In this case 
\begin{align}
r_{\mcf{J}}&=  n^{-1} \log \left \lfloor \exp \left( n r^\ddagger \right) \right \rfloor 
    \notag \\ & 
    \geq
n^{-1} \log \left( \exp \left( n r^\ddagger \right)  -1 \right)
    \\ & 
    =
r^\ddagger + n^{-1} \log \left( 1 - \exp( - n r^\ddagger) \right)
    \\ &
    \geq
r^\ddagger + n^{-1} \log \left( 1 - \exp( - n r_{\mcf{J}}) \right)
    \\ &
    \geq
r^\ddagger - n^{-1}\log \left( 2n \right)
\end{align}
where the last line is from the assumption.
Plugging in the definition of $r^\ddagger$ yields 
\begin{equation}
r_{\mcf{J}} \geq  r_{\mcf{H}} -  \frac{(1-\gamma)\ell}{4n}\lambda^2 - \frac{r_{\mcf{H}} + 2 +  \log 4n \theta}{n} 
\end{equation}
where
\begin{align}
\lambda &= \max\left( 0 ,  \min_{k \in \mcf{K}} 1 - \frac{(1+\delta)(k^2\rho_{\Delta} + \rho_{\mathrm{Dec}})}{\gamma \tau^\star(k) + (1-\gamma) \tau^\star(d_k) } \right)    
    \notag \\
\theta & = \max \left( \!1,\!\! \sqrt{3n \left[ \omega_{\mcf{K}}  + (\rho_{\Delta} + \rho_{\mathrm{Dec}}) \left(1 + \delta + 2 \lambda^2 + 2 r_{\mcf{H}} \right)\right] } \right).
    \notag
\end{align}

What remains is to prove the assumption, to that end observe
\begin{align}
&\lfloor \exp(n r^\ddagger) \rfloor 
    \notag \\ & \quad
    = 
\left \lfloor \exp \left(  (n-1) r_{\mcf{H}} -  \frac{(1-\gamma)\ell}{4}\lambda^2 - 2 -  \log 2 \theta  \right)  \right \rfloor
     \\ & \quad 
    \geq 
\left \lfloor \exp \left( \log 2n \right)  \right \rfloor  = 2n 
    \label{eq:2ndcode:2nass}
\end{align}
since
$$ (n-1) r_{\mcf{H}}   \geq \frac{(1-\gamma)\ell}{4}\lambda^2 + 2 +  \log  4n\theta .$$

~\\
\hrule
~\\
\emph{(Power)}~\\
\indent For the power constraint, 
\begin{align}
\omega_{\mcf{J}} &\leq \omega_{\mcf{L}} 
    \\
&\leq \omega_{\mcf{H}} + 2\sqrt{ 2 \omega_{\mcf{H}} \rho_{\Delta} (r_{\mcf{H}} + 1) } 
    \notag \\ & \quad 
+ \!\rho_{\Delta} \! \left(\! 1 \!+ \!(8 |\mcf{ K}|+1)  \left[ r_{\mcf{H}} \!+\!1\! + \!\frac{\log |\mcf{ K}|}{n} \right] \right)
\end{align}
since $\mbf{X}^\ddagger(m) = \mbf{X}'(m)$ whenever $m \in \{M^\ddagger\}$.
~\\
\hrule
~\\
\emph{(Average arithmetic probability of error)}~\\
\indent Next, for the average arithmetic error, let 
$$a(m) = \Pr \left( \hat m'(\mbf{X}'(m)  + \mbf{G}_{\mathrm{Dec}} ) \neq m \right)  $$
so that the average arithmetic probability of error for $M^\ddagger = \mcf{M}^\ddagger$ can be written
$$ e^{-nr^\ddagger} \sum_{m \in \mcf{M}} \idc{m}{\mcf{M}^\ddagger} a(m).    $$
From Lemma~\ref{lem:hoeffwo} though
\begin{align}
&\Pr \left( \!\!e^{-nr^\ddagger} \! \!\!\sum_{m \in \mcf{M}} \!\!\idc{m}{\{M^\ddagger\}} \! a(m) \geq \varepsilon_{\mcf{K}}(\rho_{\mathrm{Dec}}) \!+\! \!\sqrt{\frac{n}{2}e^{-nr^\ddagger}} \right) 
    \notag \\ &\quad \leq e^{-n}
\end{align}
since $e^{-nr}\sum_{m \in \mcf{M}} a(m) = \varepsilon_{\mcf{K}}(\rho_{\mathrm{Dec}}).$
Thus with probability greater than $1-e^{-n}$ the chosen $\mcf{M}^\ddagger$ will yield
\begin{align}
\varepsilon_{\mcf{J}} (\rho_{\mathrm{Dec}}) &\leq \varepsilon_{\mcf{L}}(\rho_{\mathrm{Dec}}) + \sqrt{\frac{n}{2}e^{-nr^\ddagger}} 
    \\ &\leq 
\varepsilon_{\mcf{H}}(\rho_{\mathrm{Dec}}+ \rho_{\Delta})  + \sqrt{2n e^{-nr^\ddagger}}
+ |\mcf{K}| e^{-\frac{1}{8} \ell \delta^2} \!\!.
\end{align}
\hrule
~\\
\emph{(Maximum probability of false authentication)}~\\
\indent Finally for the probability of false authentication recall from the proof of Theorem~\ref{thm:1stcode} that for each $M^\ddagger= m$, $\mbf{V}=\mbf{v}$, and attack $\mbf{Z}(\mbf{V},m) = \mbf{z}$ there exists a $\mbf{\mu} \in \mbcf{R}$ such that 
$$\mbf{Y}|\{M^\ddagger,\mbf{V},\mbf{Z} = m,\mbf{v},\mbf{z}\} = \mbf{G}_{\mbf{\tau}(m)} + \mbf{\mu}.$$
Therefore, by letting 
$$b_{m,\mbf{\mu}}(c) = \Pr\left( \hat m'(\mbf{G}_{\mbf{\tau}(m)} + \mbf{\mu}) = c \right)$$
the maximum probability of false authentication of code $\mcf{J}$ can be expressed as 
\begin{equation} \label{eq:2ndcode:mpfa}
\max_{m} \sup_{\mbf{\mu}\in \mbcf{R}} \sum_{c \in \mcf{M} \setminus \{m\}} \idc{c}{\mcf{M}^\ddagger} b_{m,\mbf{\mu}}(c) 
\end{equation}
since 
$\hat m^\ddagger(\mbf{y}) = \hat m'(\mbf{y})$ when $\hat m'(\mbf{y}) \in \mcf{M}^\ddagger$. 
An upper bound on $\alpha_{\mcf{J}}$ can therefore be obtained by computing an upper bound on $\sum_{c \in \mcf{M}-\{m\}} \idc{c}{\mcf{M}^\ddagger} b_{m,\mbf{\mu}}(c)$ that holds simultaneously for all $m \in \mcf{M}$ and $\mbf{\mu} \in \mbcf{R}.$
To that end we will employ a divide and conquer approach based on if 
$$\mbf{\mu} \in \mbcf{U}^\dagger \quad \text{where} \quad  \mcf{U}^\dagger = (-\theta,\theta) .$$

To begin, for $\mbf{\mu} \notin \mcf{U}^\dagger$ there must be a coordinate of $i\in \{1,\dots , n\}$ such that $|\mu_i| \geq \theta.$
But 
\begin{align} 
G_{\mbf{\tau}(m),i}  &\lesseqgtr  x_i(c)+t_i(c)-\mu_i \pm \sqrt{\ell (1+\delta) (f_i^2(c)\rho_{\Delta}+ \rho_{\mathrm{Dec}})}  
    \label{eq:2ndcode:pac}
\end{align}
is required for $\hat m'(\mbf{G}_{\mbf{\tau}(m)} + \mbf{\mu}) = c$ thanks to Code Modification~\ref{code:addnoise}.
Clearly the event in~\eqref{eq:2ndcode:pac} is the probability that a Gaussian random variable lies in a particular interval.
If $\mu_i \geq \theta$, then the key inequality is 
$$ G_{\mbf{\tau}(m),i}  \leq  x_i(c)+t_i(c)+ \sqrt{\ell (1+\delta) (f_i^2(c)\rho_{\Delta}+ \rho_{\mathrm{Dec}})} - \mu_i, $$
while for $\mu_i < -\theta$ it is
$$ G_{\mbf{\tau}(m),i}  \geq  x_i(c)+t_i(c)- \sqrt{\ell (1+\delta) (f_i^2(c)\rho_{\Delta}+ \rho_{\mathrm{Dec}})} - \mu_i. $$
Indeed, if $\mu_i \geq \theta$ then it follows that 
\begin{align}
    &
x_i(c) +t_i(c) + \sqrt{\ell (1+\delta)(f_i^2(c)\rho_{\Delta} + \rho_{\mathrm{Dec}})} - \mu_i
    \notag \\ & \quad 
        \leq 
\sqrt{n \omega_{\mcf{K}}} + \sqrt{\ell (1+\delta)(\rho_{\Delta} + \rho_{\mathrm{Dec}})} 
    \notag \\ & \quad \quad 
- \sqrt{3n \left[ \omega_{\mcf{K}}  + (\rho_{\Delta} + \rho_{\mathrm{Dec}}) \left(1 + \delta + 2 \lambda^2 + 2 r_{\mcf{H}} \right)\right] }
    \label{eq:2ndcode:mu1}  \\ & \quad 
   \leq 
\sqrt{n \omega_{\mcf{K}}} + \sqrt{\ell (1+\delta)(\rho_{\Delta} + \rho_{\mathrm{Dec}})} 
    \notag \\ & \quad \quad 
- \sqrt{n\omega_{\mcf{K}}} - \sqrt{n(1+\delta) (\rho_{\Delta} + \rho_{\mathrm{Dec}}) } 
    \notag \\ & \quad \quad 
- \sqrt{2n (\rho_{\Delta} + \rho_{\mathrm{Dec}}) \left[   \lambda^2 +  r_{\mcf{H}} \right]}
      \label{eq:2ndcode:mu2} \\ & \quad 
      \leq
 - \sqrt{2n (\rho_{\Delta} + \rho_{\mathrm{Dec}}) \left[   \lambda^2 +  r_{\mcf{H}} \right]}
 \label{eq:2ndcode:mu3} 
\end{align}
where~\eqref{eq:2ndcode:mu1} is plugging in the maximum values of $x_i(c)+t_i(c)$ (itself due to the power constraint), $\mu_i$, and $f_i(c);$~\eqref{eq:2ndcode:mu2} is because the concavity of the square root implies $\sqrt{3(a+b+c)} \geq \sqrt{a}+\sqrt{b}+\sqrt{c};$~\eqref{eq:2ndcode:mu3} is because $n \geq \ell.$
Hence when $\mu\geq \theta$ we can use the key inequality to bound the probability of~\eqref{eq:2ndcode:pac} as follows
\begin{align}
    &
\Phi \left( \frac{x_i(c) +t_i(c) + \sqrt{\ell (1+\delta)(f_i^2(c)\rho_{\Delta} + \rho_{\mathrm{Dec}} )} - \mu_i}{ \sqrt{\tau_i(m)}} \right)
   \notag \\ &\quad  \leq 
\Phi \left( - \sqrt{ \frac{(2 nr_{\mcf{H}} +2 n \lambda^2 ) (\rho_{\Delta}+\rho_{\mathrm{Dec}}) }{\tau_i(m)}}   \right)
    \label{eq:2ndcode:bigy1} \\ &  \quad  \leq 
\Phi \left( - \sqrt{ 2 n r_{\mcf{H}}+2 n \lambda^2}  \right)
    \label{eq:2ndcode:bigy2} \\ & \quad  \leq 
\sqrt{\frac{2}{\pi}}e^{-nr_{\mcf{H}}- n\lambda^2} 
    \label{eq:2ndcode:bigy3} 
\end{align}
where~\eqref{eq:2ndcode:bigy1} is because of~\eqref{eq:2ndcode:mu3} and $\Phi(a) > \Phi(a')$ if and only if $a > a'$;~\eqref{eq:2ndcode:bigy2} is because $0 \leq f_i^2(m) \leq 1$ for all $i$ and $m$ implies 
$$\frac{\rho_{\Delta}+\rho_{\mathrm{Dec}}}{\tau_i(m)} = \frac{\rho_{\Delta} + \rho_{\mathrm{Dec}}}{\frac{f_{i}^2(m)\rho_{\Delta} \rho_{\mathrm{Adv}} }{f^2_{i}(m)\rho_{\Delta} + \rho_{\mathrm{Adv}}} + \rho_{\mathrm{Dec}}} \geq 1;$$
finally~\eqref{eq:2ndcode:bigy3} follows from $\Phi(a) \leq \sqrt{\frac{2}{\pi}} e^{-\frac{1}{2}a^2}$ for $a\leq 0.$
A similar derivation follows for the $\mu_i \leq -\theta$ case.
Thus Equation~\eqref{eq:2ndcode:bigy3} proves
\begin{equation}
b_{m,\mbf{\mu}} \leq  \sqrt{\frac{2}{\pi}} e^{-nr_{\mcf{H}} - n \lambda^2} 
\end{equation}
for all $m \in \mcf{M}^\dagger$ and $\mbf{\mu} \notin\mbcf{U}^\dagger$, and hence
\begin{equation} \label{eq:2ndcode:nodagger}
\sum_{c \in \mcf{M}\setminus\{m\}} \idc{c}{\mcf{M}^\ddagger} b_{m,\mbf{\mu}}(c) \leq \sqrt{\frac{2}{\pi}} e^{-n \lambda^2} 
\end{equation}
for all $\mbf{\mu} \notin \mbcf{U}^\dagger$ regardless of the choice of $\mcf{M}^\ddagger.$

We now move on to the case that $\mbf{\mu} \in \mbcf{U}^\dagger$. 
Let $\mbcf{U}^\ddagger$ be the set guaranteed by Lemma~\ref{lem:onade} with respect to $\mbcf{U}^\dagger$ and positive constant $2e^{-\frac{1-\gamma}{8} \ell \lambda^2}$.
It will be important for later to note that 
\begin{equation}\label{eq:2ndcode:recall}
|\mbcf{U}^\ddagger| =  e^{n\log(2\theta) + n \frac{1-\gamma}{8} \ell \lambda^2}.
\end{equation}
From here, our strategy is to show that with high probability
\begin{equation} \label{eq:2ndcode:pfafin1}
\sum_{c \in \mcf{M}\setminus\{m\}} \idc{c}{\{M^\ddagger\}} b_{m,\mbf{\mu}}(c) \leq 2 n e^{-\frac{1}{8} \ell(1-\gamma)\lambda^2} 
\end{equation}
for all $\mbf{\mu} \in \mbcf{U}^\ddagger$ and $m \in \mcf{M}$.
With this result in hand 
\begin{equation}\label{eq:2ndcode:pfafin2}
\sum_{c \in \mcf{M}\setminus \{m\}} \hspace{-10pt} \idc{c}{\{M^\ddagger\}} b_{m,\mbf{\mu}}(c) \leq \left( 2 n  + \frac{1}{2\sqrt{n \rho_{\mathrm{Dec}}}} \right) e^{-\frac{1}{8} \ell(1-\gamma)\lambda^2}
\end{equation}
for all $\mbf{\mu} \in \mbcf{U}^\dagger$ and $m \in \mcf{M}$ follows by  Lemma~\ref{lem:onade}.
To prove~\eqref{eq:2ndcode:pfafin1} first note that for any given $\mbf{\mu} \in \mbcf{U}^\ddagger$ and $m \in \mcf{M}$ we have 
$e^{nr_{\mcf{J}}} \geq 2 n $ by Equation~\eqref{eq:2ndcode:2nass},
$$b_{m,\mbf{\mu}}(c) \leq \alpha_{\mcf{L}}^*(\rho_{\mathrm{Dec}},\rho_{\mathrm{Adv}}) \leq 2e^{-\frac{1}{8} \ell (1-\gamma)  \lambda^2}$$
by the assumptions that code $\mcf{L}$ satisfies the operational bounds set forth in Theorem~\ref{thm:1stcode}, and 
$$\sum_{c \in \mcf{M}} \frac{1}{|\mcf{M}|} e^{-nr_{\mcf{J}}} b_{m,\mbf{\mu}}(c) \leq e^{-nr_{\mcf{J}}}$$
because $\sum_{m \in \mcf{M}} b_{m,\mbf{\mu}}\leq 1.$
Hence we also have 
\begin{align}
&\Pr \left( \sum_{c \in \mcf{M}\setminus\{m\}} \idc{c}{\{M^\ddagger\}} b_{m,\mbf{\mu}}(c) \geq 2 n e^{-\frac{1}{8} \ell(1-\gamma)\lambda^2} \right)  
    \notag \\ &\quad \leq 
\exp\left(- n^2   \left[ r_{\mcf{H}} - r_{\mcf{J}} 
- \frac{(1-\gamma)\ell }{8n }\lambda^2    - \frac{3}{2n}  \right]  \right) \label{eq:2ndcode:pu}
\end{align}
by Corollary~\ref{cor:dcutting} and simple algebra. 
Using the union bound (recalling~\eqref{eq:2ndcode:recall}) and that 
$$r_{\mcf{J}}  \leq (1-n^{-1}) r_{\mcf{H}} -  \frac{(1-\gamma)\ell}{4n}\lambda^2 - \frac{2 +  \log 2 \theta}{n}$$ yields
\begin{align}
&\Pr \left(  \max_{\substack{\mbf{\mu} \in \mbcf{U}^\ddagger\\ m \in \mcf{M}}}  \sum_{c \in \mcf{M}\setminus\{m\}} \hspace{-10pt} \idc{c}{\{M^\ddagger\}} b_{m,\mbf{\mu}}(c) \!\geq\! 2 n e^{-\frac{1-\gamma}{8} \ell \lambda^2} \right)   
    \leq 
e^{-\frac{n}{2}} .\label{eq:2ndcode:pu3}
\end{align}
So, as discussed prior, Equation~\eqref{eq:2ndcode:pu3} shows that Equation~\eqref{eq:2ndcode:pfafin1} is true for all $\mbf{\mu} \in \mbcf{U}^\ddagger$ and $m \in \mcf{M}$ with exponentially high probability, hence~\eqref{eq:2ndcode:pfafin2} is true for all $\mbf{\mu} \in \mbcf{U}^\dagger$ and $m \in \mcf{M}$ with exponentially high probability as a consequence of Lemma~\ref{lem:onade}.

As a final step in the proof, we note that the upper bound on the probability of false authentication for $\mbf{\mu} \in \mbcf{U}^\dagger$, Equation ~\eqref{eq:2ndcode:pfafin2}, is greater than the upper bound on the probability of false authentication given $\mbf{\mu} \notin \mbcf{U}^\ddagger$, Equation~\eqref{eq:2ndcode:nodagger}.
Hence combining the two, we have 
\begin{align} 
\alpha_{\mcf{J}}(\rho_{\mathrm{Dec}},\rho_{\mathrm{Adv}}) &= \sup_{\substack{\mbf{\mu}\in \mbcf{R} \\ m \in \mcf{M}}} \sum_{c \in \mcf{M}\setminus\{m\}} \idc{c}{\mcf{M}^\ddagger} b_{m,\mbf{\mu}}(c) 
    \\ & \leq 
\left(2n + \frac{1}{2\sqrt{ n \rho_{\mathrm{Dec}}}}  \right)  e^{-\frac{1-\gamma}{8} \ell \lambda^2},
\end{align}
finishing the proof.








\end{IEEEproof}

\section{Theorem~\ref{thm:cap}}\label{app:cap}

\begin{IEEEproof}
First let
\begin{equation}
\rho_{\Delta} = o(1) \quad \text{ and } \quad \delta = o(\rho_{\Delta})
\end{equation}
be sufficiently large and note by the channel capacity theorem of Shannon~\cite{shannon1948mathematical}, for all positive real finite numbers $a$ and $b$ there exists a sequence of codes $\mcf{H}_{(n)} = (\mbf{x}_{(n)}: \mcf{M}_{(n)} \rightarrow \mbcf{R}_{(n)},\hat m_{(n)} : \mbcf{R}_{(n)} \rightarrow \mcf{M}_{(n)})$ such that 
\begin{align}
\lim_{n \rightarrow \infty} r_{\mcf{H}_{(n)}} &= \frac{1}{2} \log \left( 1 + \frac{a}{b} \right) 
    \notag \\ 
\lim_{n \rightarrow \infty}  \omega_{\mcf{H}_{(n)}}  &= a 
    \notag \\ 
\varepsilon_{\mcf{H}_{(n)}}(b) &=  0.
    \notag 
\end{align}
Letting $a = \rho - O(\sqrt{\rho_{\Delta}})$ and $b = \rho_{\Delta} + \rho_{\mathrm{Dec}}$ and applying Theorem~\ref{thm:2ndcode} yields a sequence of codes $\mcf{J}_{n}$ such that  
\begin{align}
\lim_{n\rightarrow \infty} \omega_{\mcf{J}_{(n)}}  &\leq \lim_{n\rightarrow \infty}  \omega_{\mcf{H}_{(n)}} + O( \sqrt{\rho_{\Delta}} ) 
    =
\rho 
    \\ 
\lim_{n\rightarrow \infty}  \varepsilon_{\mcf{J}_{(m)}}(\rho_{\mathrm{Dec}} ) &\leq  \lim_{n\rightarrow \infty}  \varepsilon_{\mcf{H}}(\rho_{\mathrm{Dec}}+ \rho_{\Delta})  + e^{-O(n \delta^2)}
     \\ &
     =
0
    \\
\lim_{n\rightarrow \infty}  \alpha_{\mcf{J}_{(n)}}(\rho_{\mathrm{Dec}},\rho_{\mathrm{Adv}} ) &\leq  e^{-O(n \rho_{\Delta}^2)}.
    =
0,
\end{align}
while
\begin{align}
&\lim_{n\rightarrow \infty} r_{\mcf{J}_{(n)}} 
    \notag \\ &
    \geq  
\lim_{n \rightarrow \infty} r_{\mcf{H}_{(n)}} - O\left(\rho_{\Delta}^2+\frac{\log n}{n} \right)
    \\&
    = 
 \lim_{n \rightarrow \infty} \frac{1}{2} \log \left( 1 + \frac{\rho- O(\sqrt{\rho_{\Delta}})}{\rho_{\Delta}+ \rho_{\mathrm{Dec}}} \right) - O\left(\rho_{\Delta}^2+\frac{\log n}{n} \right) 
    \\&
    =
\frac{1}{2} \log \left( 1 + \frac{\rho}{\rho_{\mathrm{Dec}}} \right) 
\end{align}
This proves $\frac{1}{2} \log \left( 1 + \frac{\rho}{\rho_{\mathrm{Dec}}} \right)$ is achievable, and by~\cite{shannon1948mathematical} it is also an upper bound, hence 
\begin{align}
c(\rho,\rho_{\mathrm{Dec}},\rho_{\mathrm{Adv}}) = \frac{1}{2} \log \left( 1 + \frac{\rho}{\rho_{\mathrm{Dec}}} \right) 
\end{align}
for $\rho_{\mathrm{Adv}}>0.$

On the other hand if $\rho_{\mathrm{Adv}}=0,$ then $\mbf{V} = \mbf{X}(M)$ for any code $\mbf{X}$, $\hat m$. 
Hence the adversary may choose $\mbf{Z}(\mbf{V},M) = \mbf{X}(m') - \mbf{V},$ to produce 
$$\mbf{Y} = \mbf{X}(m') + \mbf{G}_{\mathrm{Dec}}.$$
Thus
$$\alpha_{\mbf{X},\hat m}(\rho_{\mathrm{Dec}},0) \geq 1 -\varepsilon_{\mbf{X},\hat m}(\rho_{\mathrm{Dec}},0),$$
and consequently
\begin{equation}
c(\rho,\rho_{\mathrm{Dec}},0) = 0 .
\end{equation}

\end{IEEEproof}

\end{document}